\documentclass[12pt,amsmath,amssymb,nofootinbib]{revtex4}
\usepackage{amsmath,amsfonts,amsthm,amscd,amssymb,latexsym}
\usepackage{bm}
\usepackage{dcolumn}
\usepackage{graphicx}
\usepackage{epstopdf}
\usepackage{color}
\usepackage{epsf}
\usepackage{epsfig}
\usepackage{graphicx, epic, eepic, color}
\usepackage{soul}

\newcommand{\beq}{\begin{equation}}
\newcommand{\eeq}{\end{equation}}
\newcommand{\bea}{\begin{eqnarray}}
\newcommand{\eea}{\end{eqnarray}}

\begin{document}

\title{Static and Dynamic Melvin Universes}

\author{Donato Bini$^{1,2}$ and Bahram Mashhoon$^{3,4}$}
  \affiliation{
$^1$Istituto per le Applicazioni del Calcolo ``M. Picone" CNR, I-00185 Rome, Italy\\
$^2$INFN, Sezione di Roma Tre, I-00146 Rome, Italy\\
$^3$ School of Astronomy,
Institute for Research in Fundamental Sciences (IPM),
P. O. Box 19395-5531, Tehran, Iran\\
$^4$Department of Physics and Astronomy,
University of Missouri, Columbia,
Missouri 65211, USA\\
}

\date{\today}
\begin{abstract}
We briefly review the known properties of Melvin's magnetic universe and study the propagation of test charged matter waves in this static spacetime. Moreover, the possible correspondence between the wave perturbations on the background Melvin universe and the motion of charged test particles is discussed. Next, we explore a simple scenario for turning Melvin's static universe into one that undergoes gravitational collapse.  In the resulting dynamic  gravitational field, the formation of cosmic double-jet configurations is emphasized. 
\end{abstract}

\maketitle

\section{Introduction}

The Melvin magnetic universe~\cite{Bonnor:1954,Melvin} is an exact static solution of Einstein-Maxwell equations in which Maxwell's magnetic pressure is balanced by gravitational attraction. In this cylindrically symmetric spacetime, the magnetic field is parallel to the regular axis of symmetry. The stability as well as other essential properties of Melvin's magnetic universe have been extensively studied by a number of authors~\cite{Melvin:1965zza, Thorne:1965, MelWal, Lim:2020fnx}. The spacetime metric is given by~\cite{Stephani:2003tm, GrPo, BSA} 
\beq \label{I1}
ds^2= \left(1+\frac{GB_0^2\rho^2}{4c^4}\right)^2(- c^2dt^2+d\rho^2+dz^2)+\frac{\rho^2}{\left(1+\frac{GB_0^2\rho^2}{4c^4}\right)^2}d\phi^2\,,
\eeq
where $B_0 > 0$ is a constant.  Let us introduce a constant parameter $a$ that has the dimensions of length via
\beq
\label{I2}
 a := \frac{2c^2}{G^{1/2} B_0}\,.
\eeq
It will turn out that $B_0$ has the interpretation of the measured magnitude of the magnetic field strength at the axis of cylindrical symmetry. The magnetic field strength monotonically decreases away from the symmetry axis; therefore, $B_0$ is the maximum magnetic field strength in the Melvin universe.  The average magnitude of Earth's magnetic field at its surface is about $0.5$ Gauss; for $B_0 = 0.5$ Gauss, $a \approx 5$ Mpc.  Henceforth, we use the convention that the field is oriented along the positive $z$ axis  and employ units such the $G = c = 1$, unless specified otherwise.

 Consider the transformation to dimensionless spacetime coordinates given by
\beq
\label{I3}
(t, \rho, z ) \mapsto (at, a \rho, az)\,.
\eeq
Using the new dimensionless coordinates, the spacetime metric takes the form
\beq
\label{I4}
ds^2=a^2 \left[(1+\rho^2)^2(-dt^2+d\rho^2+dz^2)+\frac{\rho^2}{(1+\rho^2)^2}d\phi^2 \right]\,,
\eeq
where $a$ is the length scale that henceforth  characterizes  the Melvin universe.

Einstein's gravitational field equation can be written as 
\beq
\label{I5}
R^{\mu}{}_{\nu} - \frac{1}{2} \delta^\mu_\nu R = \frac{8 \pi G}{c^4} T^{\mu}{}_{\nu}\,,
\eeq
where the traceless electromagnetic energy-momentum tensor is given by
\beq
\label{I6}
 4 \pi \, T^{\mu}{}_{\nu} = F^{\mu}{}_{\alpha}F_{\nu}{}^{\alpha}  - \frac{1}{4} \delta^\mu_\nu \,F_{\alpha \beta}F^{\alpha \beta}\,.
\eeq
We assume throughout that the cosmological constant vanishes. The Melvin universe in the presence of a cosmological constant has been considered in Refs.~\cite{Astorino:2012zm, Zofka:2019yfa}.

For the simple case of metric~\eqref{I4}, $ds^2 = g_{\mu \nu} dx^\mu dx^\nu$, the nonvanishing coordinate components of the Riemann tensor can be obtained from
\bea \label{I7}
R_{0101}&=& 2 a^2(1 -\rho^2)=-R_{1313}\,, \nonumber\\ 
R_{0202}&=& 2 a^2 \rho^2\frac{(1-\rho^2)}{(1+\rho^2)^4}=-R_{2323}\,, \nonumber  \\
R_{0303}&=& 4 a^2\rho^2\,,\qquad R_{1212}= 4 a^2 \rho^2 \frac{(2-\rho^2)}{(1+\rho^2)^4}\,.
\eea
The Ricci tensor $R_{\alpha\beta}=R^\mu{}_{\alpha\mu\beta}$ is diagonal and has coordinate components
\beq \label{I8}
R^\alpha{}_\beta=\frac{4}{a^2(1+\rho^2)^4}\,{\rm diag}[-1,1,1,-1]\,.
\eeq 

At the same time, metric~\eqref{I4} represents a static electrovacuum spacetime with electromagnetic potential
\beq  \label{I9}
A= - \frac{a}{1+\rho^2} d\phi\,,\qquad  A\cdot A=\frac{1}{\rho^2}\,
\eeq 
and associated electromagnetic field tensor $F=dA$ given by
\beq   \label{I10}
F= F_{\rho \phi}\,d\rho \wedge d\phi\,, \qquad F_{\rho \phi} = \frac{2a\rho }{(1+\rho^2)^2}\,.
\eeq
In this case, $A_{\alpha; \beta}$ is antisymmetric, which implies $A_{(\alpha; \beta)}=0$  and $A^\alpha{}_{;\alpha}=0$. Moreover, the two algebraic electromagnetic invariants are
\beq    \label{I11}
I_1=\frac12 F_{\alpha\beta}F^{\alpha\beta} := B^2 =\frac{4}{a^2(1+\rho^2)^4}\,,\qquad I_2=\frac12 F^*_{\alpha\beta}F^{\alpha\beta}=0\,.
\eeq
Similarly, the Kretschmann invariant is
\beq  \label{I12}
R_{\alpha\beta\gamma\delta}R^{\alpha\beta\gamma\delta}=C_{\alpha\beta\gamma\delta}C^{\alpha\beta\gamma\delta}+ 2 R_{\alpha\beta}R^{\alpha\beta}-\frac{1}{3} R^2\,.
\eeq
Melvin's universe is of Petrov type D with
\beq  \label{I13}
R_{\alpha\beta\gamma\delta}R^{\alpha\beta\gamma\delta}=\frac{64(3\rho^4-6\rho^2+5)}{a^4(1+\rho^2)^8}\,,\quad 
R^*_{\alpha\beta\gamma\delta}R^{\alpha\beta\gamma\delta}=0\,
\eeq
and the Weyl conformal tensor invariant
\beq  \label{I14}
 C_{\alpha\beta\gamma\delta}C^{\alpha\beta\gamma\delta}=\frac{192(\rho^2-1)^2}{a^4(1+\rho^2)^8}\,.
\eeq

Consider the family of static reference observers in the Melvin universe that are all spatially at rest. Each such fiducial observer carries a natural orthonormal tetrad frame $e_{\hat \alpha}$, 
\beq
\label{I15}
e_{\hat 0}=\frac{1}{a(1+\rho^2)}\partial_t\,,\qquad e_{\hat 1}=\frac{1}{a(1+\rho^2)}\partial_\rho\,,\qquad
e_{\hat 2}=\frac{1+\rho^2}{a\rho}\partial_\phi\,,\qquad e_{\hat 3}=\frac{1}{a(1+\rho^2)}\partial_z\,,
\eeq
adapted to its world line with proper time $\tau$, unit timelike tangent vector $e_{\hat 0}$ and spatial frame that is directed along the background cylindrical coordinate axes. The fiducial observer carries the tetrad frame along its world line according to 
\begin{equation}\label{I16}
\frac{D e^{\mu}{}_{\hat \alpha}}{d\tau} = \mathbb{F}_{\hat \alpha}{}^{\hat \beta} \,e^{\mu}{}_{\hat \beta}\,,
\end{equation}
where $\mathbb{F}_{\hat \alpha \hat \beta}$ is the fiducial observer's acceleration tensor that is antisymmetric due to the tetrad orthonormality condition, namely, 
\begin{equation}\label{I17}
g_{\mu \nu} \,e^\mu{}_{\hat \alpha}\,e^\nu{}_{\hat \beta}= \eta_{\hat \alpha \hat \beta}\,.
\end{equation}

The acceleration tensor can be naturally decomposed into its ``electric" and ``magnetic" parts, namely, $\mathbb{F}_{\hat \alpha \hat \beta} \mapsto (-\mathbf{g}, \boldsymbol{\omega})$, where $\mathbf{g}(\tau)$ and $\boldsymbol{\omega}(\tau)$ are spacetime scalars  that represent the translational and rotational accelerations  of the fiducial observer, respectively. The deviation of reference observer's world line  from a geodesic is measured by $\mathbf{g}$, while $\boldsymbol{\omega}$ is the angular velocity of the rotation of the reference observer's spatial frame with respect to a locally nonrotating (i.e. Fermi-Walker transported) frame.  The reference observer's acceleration tensor follows from the equation of motion of the frame~\eqref{I16} and has components
\beq    \label{I18}
\mathbf{g}=\frac{2\rho}{a(1+\rho^2)^2}\,e_{\hat1}\,,\qquad {\boldsymbol \omega}=0\,.
\eeq
Indeed, the  world lines of the family of fiducial observers  form a congruence with acceleration along the outward radial direction in order to compensate for the inward gravitational attraction, while both the expansion and vorticity of the congruence vanish identically.

Projecting the electromagnetic field tensor on fiducial tetrad~\eqref{I15} leads to the 
nonvanishing measured component of the magnetic field
\beq    \label{I19}
F_{\hat1 \hat 2}= - F_{\hat2 \hat 1} = \frac{2}{a(1+\rho^2)^2} := B\,, \qquad B = \frac{B_0}{(1+\rho^2)^2}\,.
\eeq
The measured magnetic field $B$ monotonically decreases always from the axis of cylindrical symmetry and vanishes at infinity. A similar analysis of the nature of the Melvin gravitational field as determined locally by the fiducial observers is contained in Appendix A via the establishment of a Fermi normal coordinate system. 

In the next two sections, we briefly discuss the motion of neutral and charged test particles in the Melvin magnetic universe. In this connection, see Refs.~\cite{Melvin:1965zza, Thorne:1965} for previous work regarding geodesics of the Melvin universe; furthermore,  the motion of charged particles has been discussed  in Refs.~\cite{MelWal, Lim:2020fnx}.

\section{Timelike and Null Geodesics}

Consider the motion of a free neutral test particle in the Melvin universe and let  $x^\mu(\tau)$ be its future-directed timelike geodesic world line, where $\tau$ is its proper time. The Melvin universe has Killing vectors ($\partial_t, \partial_\phi, \partial_z$), due to its invariance under translations in $(t, \phi, z)$, and  $z\,\partial_t + t\,\partial_z$ due to its invariance under Lorentz boosts in the $(t, z)$ plane. Projecting $u^\mu = dx^\mu/d\tau$, the unit four-velocity vector of the free particle, upon the Killing vectors, we find the canonical momenta (per unit inertial mass of the test particle) that are constants of geodesic motion, namely, 
\beq    \label{T1}
a^2 (1+\rho^2)^2 \frac{dt}{d\tau} = P_t\,, \quad  \frac{a^2\rho^2}{(1+\rho^2)^2} \frac{d\phi}{d\tau} = P_\phi\,, \quad  a^2 (1+\rho^2)^2 \frac{dz}{d\tau} = P_z\,,
\eeq
while the constant of the motion, $-zP_t + tP_z$, corresponding to projection of $u^\mu$ upon the boost Killing vector simply implies that $z$ varies linearly with $t$, which means that  the particle  in general  has uniform linear motion parallel to the axis of symmetry in terms of coordinate time $t$, a circumstance that is already contained in Eq.~\eqref{T1}. That is,
\beq     \label{T2}
\frac{dz}{dt}=\frac{P_z}{P_t}\,,\qquad z(t)-z(0)=\frac{P_z}{P_t} \,t\,.
\eeq
Here, per unit mass of the free test particle, $P_t >0$ has the interpretation of its energy, while $P_\phi$ is its orbital angular momentum about the axis of cylindrical symmetry and $P_z$ is its momentum along this axis.  

From $u^\mu u_\mu = -1$, we find
\beq    \label{T3}
a^4 (1+\rho^2)^4 \left(\frac{d\rho}{d\tau}\right)^2  + V(\rho) = P_t^2-P_z^2\,, \quad V(\rho) = a^2 (1+\rho^2)^2 +  \frac{P_\phi^2(1+\rho^2)^4}{\rho^2}\,,
\eeq
which implies that timelike geodesics exist provided $\mathbb{E} := (P_t^2 - P_z^2)^{1/2} \ge \mathbb{E}_0$.  Here, $\mathbb{E}_0>0$ is a constant such that  $\mathbb{E}_0^2 = V(\rho_0)$, where 
$\rho_0$ is  the minimum of the  manifestly positive effective potential $V(\rho)$. For $P_\phi \ne 0$, the effective potential exhibits  a centrifugal barrier for $\rho \to 0$, monotonically decreases to its minimum at $\rho_0$ and then monotonically increases to infinity. Indeed, $V(\rho) \sim \rho^6$ as $\rho \to \infty$. It is important to note that these features belong to the dominant term in the potential that is proportional to $P_\phi^2$ and has its minimum at $1/\sqrt{3}$. Hence, $V(\rho)$ is a potential well that is infinitely deep and there are in general two turning points $\rho_1$ and $\rho_2$, $\rho_1 < \rho_0 < \rho_2$. For $P_z \ne 0$, the  motion is in general helical, while it is elliptic in the  plane orthogonal to the symmetry axis and the particle can never reach $\rho = \infty$.  Moreover, the stable circular orbits with radius $\rho_0$ are such that
\beq    \label{T4}
\mathbb{E}_0 = a (1+\rho_0^2) \left(\frac{1-\rho_0^2}{1-3\rho_0^2}\right)^{1/2}\,, \quad (P_\phi)_0^2 =  \mathbb{E}_0^2 \frac{2\rho_0^4}{(1-\rho_0^2)(1+\rho_0^2)^4}\,
\eeq 
and
\beq    \label{T5}
 \left(\frac{d\phi}{d\tau}\right)_0^2 =  \frac{2(1+\rho_0^2)^2}{a^2(1-3\rho_0^2)}\,. 
\eeq 
Thus timelike circular geodesic orbits exist for $\rho_0 < 1/\sqrt{3}$. 

We should mention the special case of geodesic motion involving motion at constant $\phi$.  The effective potential is then simply given by $V(\rho) = a^2 (1+\rho^2)^2$ and we must have $\mathbb{E}_0 \ge a$. For $\mathbb{E}_0 > a$, the motion is oscillatory along the radial direction and the radial turning point is determined by $(-1 + \mathbb{E}_0/a)^{1/2}$; in this case, the equation of motion~\eqref{T3} can be integrated using elliptic integrals. On the other hand, this case can degenerate into motion purely parallel to the symmetry axis at constant $\rho = (-1 + \mathbb{E}_0/a)^{1/2}$. Finally, 
for $\mathbb{E}_0 = a$, the particle stays on the axis of symmetry $\rho = 0$ and can only move along this axis; in fact, this case can be regarded as the degenerate limit of circular geodesic orbits as they approach the axis of cylindrical symmetry.    

Imagine now a null geodesic with tangent vector $k^\mu = d x^\mu/d\eta$, where $\eta$ is an affine parameter along its world line. The transition from a timelike to null geodesic can be effected through the relation  $d\tau = m\, d\eta$. As the test particle's mass tends to zero,  $d\tau \to 0$  such that $k^\mu := m\, u^\mu$ becomes a null vector, i.e.  $k^\mu k_\mu \to 0$.  The treatment of null geodesic orbits is then much the same as above with $(P_t, P_\phi, P_z) m = (p_t, p_\phi, p_z)$, except that the first term in the effective potential, $a^2 (1+\rho^2)^2$,  now disappears and the remaining centrifugal term has a minimum at 
$(\rho_0)_{\rm null} = 1/\sqrt{3}$, which is the radius of the unique null circular geodesic orbit with $d\phi/dt= \pm 16\sqrt{3}/9$ for constant $z$.   For the special case of constant $\phi$, the effective potential completely disappears with $p_\phi = 0$; hence, a null geodesic traveling outward along the radial direction can  reach radial infinity.

Finally, let us mention that timelike circular geodesic orbits measure the strength of the attraction of gravity and the distribution of these orbits in this cylindrically symmetric case indicates that the attraction of gravity increases away from the symmetry axis and reaches a maximum at $(\rho_0)_{\rm null} = 1/\sqrt{3}$. This is indeed the case, as illustrated by the nongravitational  radial acceleration~\eqref{I18} experienced by observers at rest in the Melvin universe. This radial acceleration, $g = (2\rho/a)(1+\rho^2)^{-2}$, counteracts the attraction of gravity; that is, $g$ is the magnitude of the strength of the attraction of gravity as experienced by  observers that are spatially at rest. We plot $a g$  versus $\rho$ in Figure 1. As illustrated in this figure, the attraction of gravity increases monotonically with radius $\rho$ and has a maximum, $g_{\rm max} \approx 0.65/a$,  at $1/\sqrt{3}$; beyond this point, no  timelike or null circular geodesic orbits are allowed.  Figure 1 demonstrates another peculiar feature of the Melvin universe: We note that below $g_{\rm max}$, the same radial attraction of gravity is experienced at two different radii, one near the axis of symmetry, $\rho_{\rm near} < 1/\sqrt{3}$, and one far from this axis, $\rho_{\rm far} > 1/\sqrt{3}$. On the other hand, the magnetic energy density as measured by the observers at rest, $B^2/(8 \pi)$, monotonically decreases away from the symmetry axis and vanishes as $\rho \to \infty$.   
\begin{figure}
\includegraphics[scale=0.4]{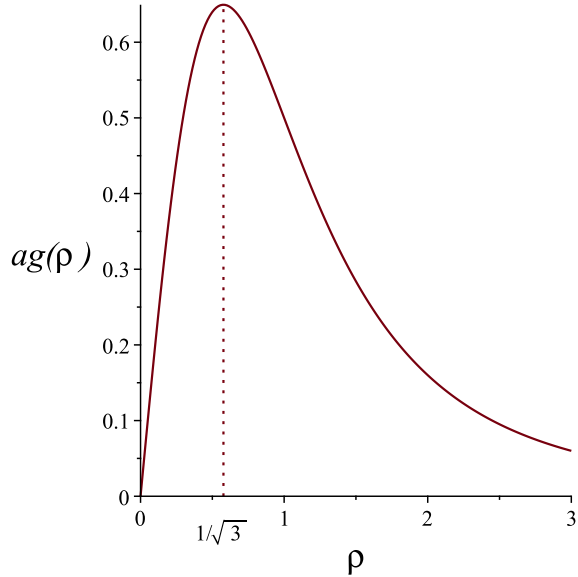}
\caption{Plot of $a\,g(\rho) = 2 \rho/(1+\rho^2)^2$ versus $\rho$, which indicates how the attraction of gravity increases with $\rho$ until $(\rho_0)_{\rm null} = 1/\sqrt{3}$. Timelike circular geodesic orbits occur in the Melvin magnetic universe for $0 < \rho_0 < 1/\sqrt{3}$, while $(\rho_0)_{\rm null} = 1/\sqrt{3}$.
}
\end{figure}

\section{Motion of charged particles}

Let us consider a test particle of mass $m$ and electric charge $q$ moving along a future-directed timelike world line $x^\alpha=x^\alpha(\tau)$ parametrized by proper time $\tau$.  Neglecting radiative effects throughout, the equations of motion read
\beq
\label{M1}
\frac{DU^\mu}{d\tau} \equiv \frac{dU^\mu}{d\tau}+\Gamma^\mu{}_{\alpha\beta}U^\alpha U^\beta= \mathcal{A}^\mu \,, \qquad    \mathcal{A}^\mu = \frac{q}{m} F^\mu{}_{\nu}U^\nu\,,
\eeq
where $U^\alpha= dx^\alpha /d\tau$ is the 4-velocity and $\mathcal{A}^\mu$ is the 4-acceleration of the particle.  Let $K^\mu$ be a Killing vector field and consider 
\beq
\label{M2}
\frac{d(U^\mu K_\mu)}{d\tau}= \frac{DU^\mu}{d\tau} K_\mu + K_{(\mu;\nu)} U^\mu U^\nu = \mathcal{A}^\mu K_\mu\,.
\eeq
In the present case, the 4-acceleration is given by the Lorentz force per unit mass of the particle, namely, $\mathcal{A}_\mu = \hat{q} F_{\mu \nu}U^\nu$, where 
\beq
\label{M3}
\hat{q} := \frac{q}{m}\,
\eeq
is the electric charge per unit mass of the particle. Let us recall that in the Melvin universe, the only nonzero component of the electromagnetic field tensor is $F_{\rho \phi} = - F_{\phi \rho} = 2 a \rho/(1+\rho^2)^2$. Therefore, Eq.~\eqref{M2} implies
\beq    \label{M4}
a^2 (1+\rho^2)^2 \frac{dt}{d\tau} = \hat{P}_t\,,  \qquad  a^2 (1+\rho^2)^2 \frac{dz}{d\tau} = \hat{P}_z\,,
\eeq
and
\beq    \label{M5}
\frac{d}{d\tau}\left[\frac{a^2\rho^2}{(1+\rho^2)^2} \frac{d\phi}{d\tau}\right] = -\hat{q} F_{\rho \phi} \frac{d\rho}{d\tau}\,,
\eeq
which can be integrated and we find
\beq    \label{M6}
\frac{a^2\rho^2}{(1+\rho^2)^2} \frac{d\phi}{d\tau} -\hat{q} \frac{a}{1+\rho^2} = \hat{P}_\phi\,.
\eeq
The angular momentum of the particle per unit mass is the sum of the orbital part plus an additional piece, i.e. $\hat{q} A_\phi$,  due to the interaction of the charged particle with the magnetic field.  It follows from Eq.~\eqref{M4} that, as in Eq.~\eqref{T2}, $z$ is in general a linear function of $t$ and the motion is helical for $ \hat{P}_z \ne 0$. 

The 4-velocity $U^\mu$ is a unit timelike vector; hence, $U^\mu U_\mu = -1$. This implies, in view of Eqs.~\eqref{M4} and~\eqref{M6}, that 
\beq    \label{M7}
a^4 (1+\rho^2)^4 \left(\frac{d\rho}{d\tau}\right)^2  + \mathcal{V}(\rho) = \hat{P}_t^2- \hat{P}_z^2\,, \quad \mathcal{V}(\rho) = a^2 (1+\rho^2)^2 +  \frac{(1+\rho^2)^4}{\rho^2}\left[\hat{P}_\phi + \hat{q} \frac{a}{1+\rho^2}\right]^2\,,
\eeq
which is the natural generalization of Eq.~\eqref{T2} to the case of charged particle motion. The effective potential $\mathcal{V}$ for charged particles is manifestly positive as well; therefore, we must have 
$\hat{P}_t^2 > \hat{P}_z^2$.  The general graph of $\mathcal{V}$ versus $\rho$ is qualitatively the same as before; that is, for $\rho: 0\to \infty$, $\mathcal{V}$ monotonically decreases  from $\infty$, has a minimum at $\hat{\rho}_0$ and  increases  monotonically to $\infty$. The general motion of charged particles is helical; moreover, in the 
$(x, y)$ plane with $x = \rho \cos \phi$ and $y = \rho \sin \phi$, we have two turning points corresponding to ellipses for $\hat{\mathbb{E}} := (\hat{P}_t^2 - \hat{P}_z^2)^{1/2} > \hat{\mathbb{E}}_0 > 0$. Here,  
$\mathcal{V}(\hat{\rho}_0) = \hat{\mathbb{E}}_0^2$ for stable timelike circular orbits that correspond to the minimum of the effective potential. Let $\varpi =  \hat{\rho}_0^2$; then, we have a cubic equation 
for $\varpi$, namely, 
\beq    \label{M8}
3 \hat{P}_\phi^2\,\varpi^3 + ( 5 \hat{P}_\phi^2 + 4 \hat{P}_\phi \hat{q} a + 2a^2) \varpi^2 = (\hat{P}_\phi  + \hat{q} a)^2 (1-\varpi)\,.
\eeq
This cubic equation has only one real solution for $\varpi$ that is positive, which can be simply demonstrated via the intersection of the graphs of both sides of Eq.~\eqref{M8}.  A detailed discussion of circular orbits is contained in Appendix B, where it is demonstrated that
\beq
\label{M9}
\hat{\rho}_0 < \frac{1}{\sqrt{3}} \left( 1 +\frac{1}{2} \hat{q}^2\right)\,, 
\eeq
which agrees with the neutral case for $q=0$. 

Let us note that the dominant terms in the effective potential near $\rho = 0$ and $\rho =\infty$ are
\beq
\label{M10}
\mathcal{V}(\rho)\bigg|_{\rho\to 0 }\sim  \frac{(\hat P_\phi+a\hat q)^2}{\rho^2}\,, \qquad  \mathcal{V}(\rho)\bigg|_{\rho\to \infty } \sim  \hat P_\phi^2\,\rho^6\,;
\eeq
hence, we need to consider special cases given by $\hat P_\phi + a \hat q = 0$ and $\hat P_\phi = 0$. 

For $\hat P_\phi + a \hat q = 0$, the centrifugal barrier disappears and the effective potential starts from $a^2$ at $\rho=0$ and monotonically increases to infinity. For $\hat{\mathbb{E}} > a$,  the effective potential has a turning point and the charged particle thus oscillates radially back and forth crossing the axis of symmetry and going out to a certain finite radial distance, while the azimuthal motion is given by $d\phi/d\tau = - (\hat{q}/a)(1+\rho^2)$.  For $\hat{\mathbb{E}} = a$, the charged particle remains at the symmetry axis; in some sense, this is the limit of timelike circular orbits in this case. 

Finally, for $\hat P_\phi = 0$, there is a minimum in the effective potential determined solely by $\hat q$ and the helical motion is qualitatively the same as before with 
$d\phi/d\tau = (\hat{q}/a)(1+ 1/\rho^2)$; this special case is illustrated in Figure 2.

\begin{figure}
\[
\begin{array}{cc}
\includegraphics[scale=0.4]{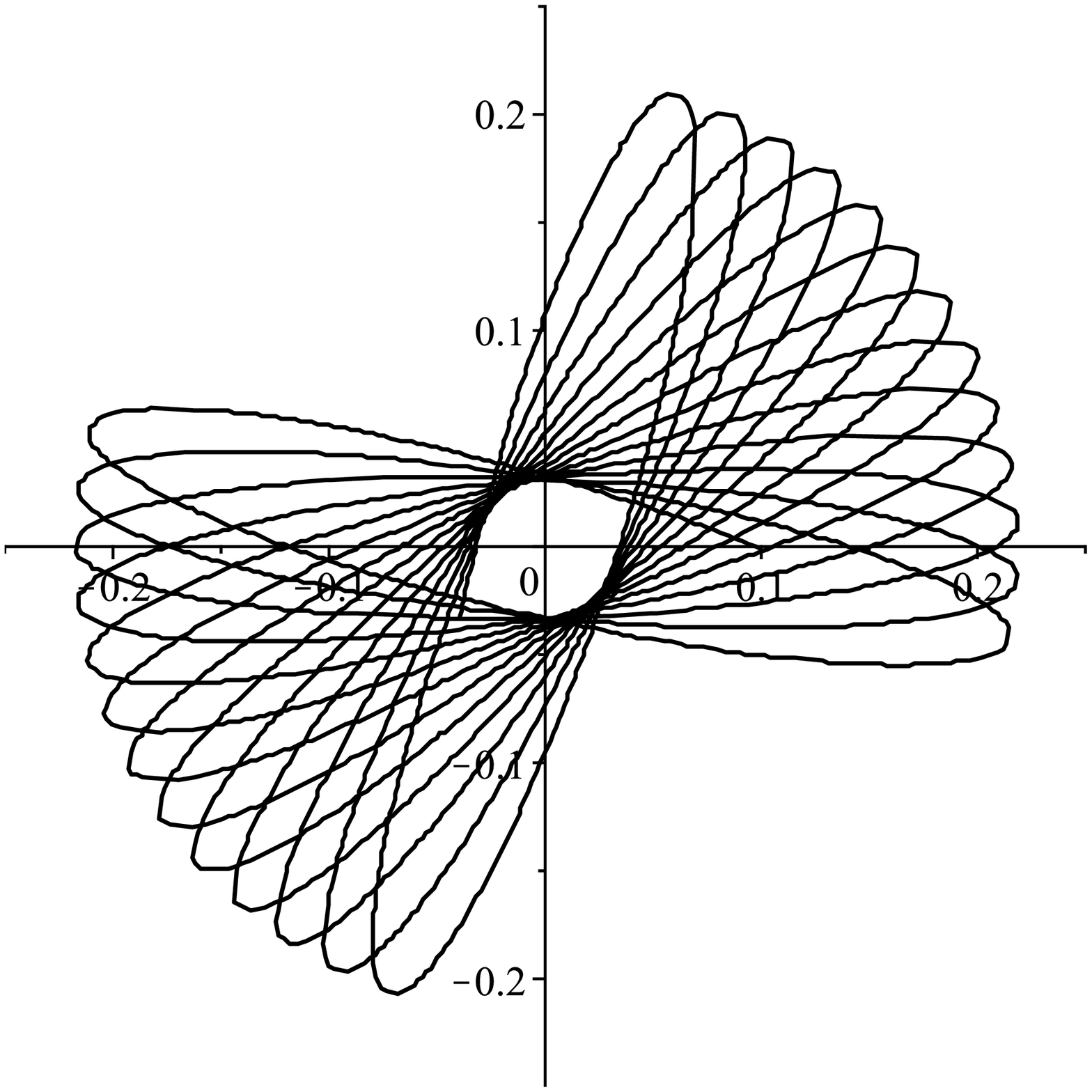} & \includegraphics[scale=0.4]{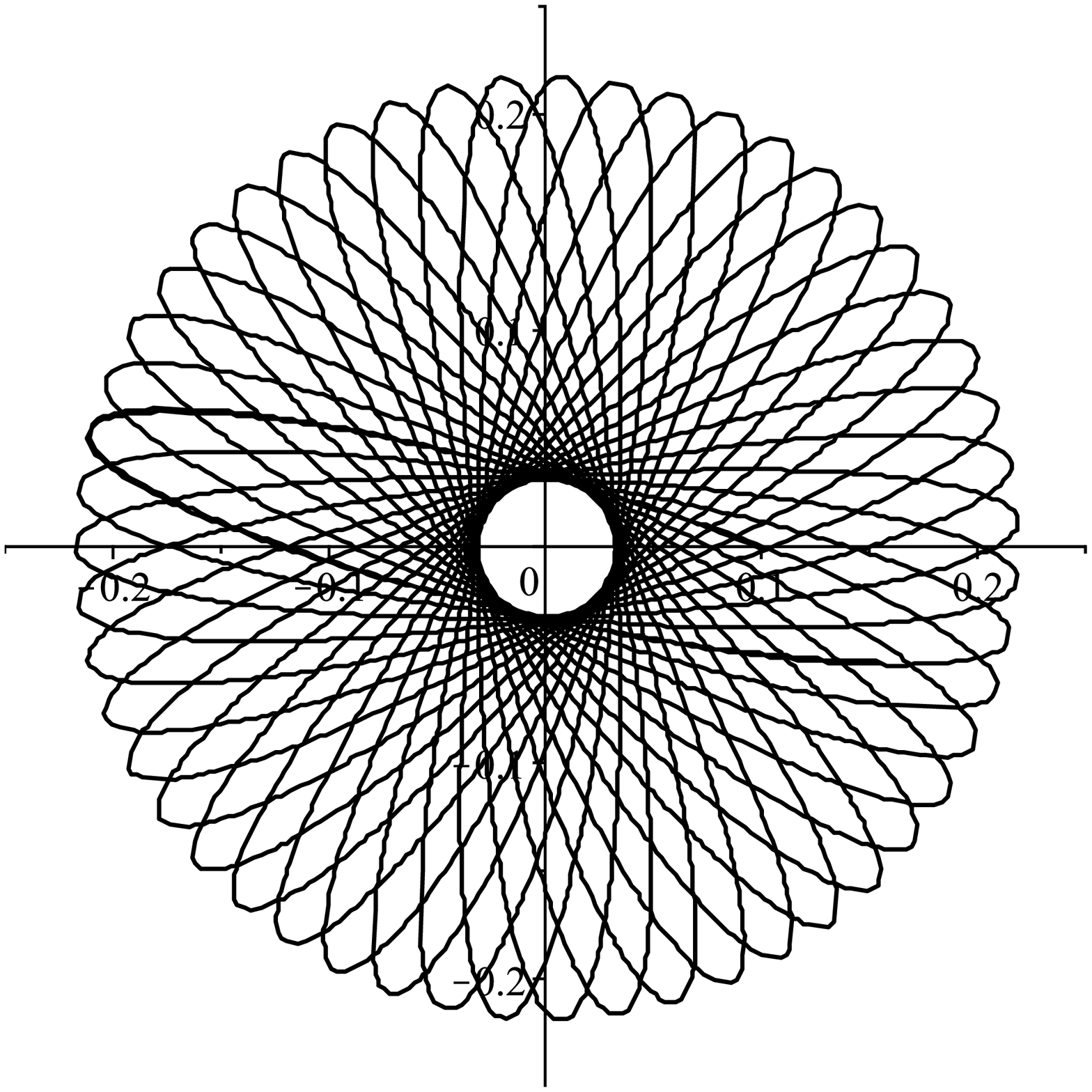}\cr
(a) & (b) \cr
\end{array}
\]
\caption{The rosette form of the orbit of a charged particle in the $(x, y)$ plane, where $x = \rho \cos \phi$ and $y = \rho \sin \phi$. The numerical integration involves parameters 
$\hat P_\phi=0$, $\hat{P}_t\approx0.105$, $\hat P_z = 0$, $q/m=0.01$ and $a=0.1$  as well as initial conditions at proper time $\tau = 0$ given by
$t(0)=0$, $\rho(0)=0.1$, $\phi(0)=0$, $z(0) = 0$ and $(d\rho/d\tau)(0) \approx-2.588$. The effective potential has two turning points  in this case:  the minimum radius is $\rho_{\rm min}\approx0.032$, while the maximum radius is $\rho_{\rm max}\approx 0.218$. In panel (a), the integration time is $\tau\in  [0,6]$, while in panel (b), $\tau\in  [0,12]$. Units are not specified here, since the plots are purely for illustration.
}
\end{figure}


We must digress here and mention that most astronomical objects are on average neutral; indeed, the total charge-to-mass ratio for an astronomical body can be expressed in terms of the dimensionless quantity $\hat{q} = q/ (\sqrt{G} m)$ and in general $|\hat{q}| \ll 1$~\cite{MaPa}. On the other hand, this ratio for a proton  is of order $10^{18}$, since the elementary charge is $\approx 4.8 \times 10^{-10}$ esu and the mass of the proton is $\approx 1.67 \times 10^{-24}$ gm.  We work primarily within the domain of classical physics in this paper; therefore, we assume that electrically charged classical test particles carry charges such that  $|\hat{q}| \lesssim 1$.

To discuss the transition to null paths, we consider the relation $d \tau = m \,d\eta$ such that as $m \to 0$, the future-directed timelike path of the charged test particle approaches a null world line given by
\beq
\label{M11}
\frac{d^2x^\mu}{d\eta^2}+\Gamma^\mu{}_{\alpha\beta}\frac{dx^\alpha}{d\eta} \frac{dx^\beta}{d\eta}= q F^\mu{}_{\nu}\frac{dx^\nu}{d\eta}\,,
\eeq
where $\eta$ is an affine parameter along the null path. At this level, classical general relativity appears to allow the  possibility of existence of electrically charged massless particles. Such particles do not exist, however, in the real (quantum) world. Therefore, we set $q=0$ in Eq.~\eqref{M11} due to the nonexistence of massless charged particles. In the transition to a null path, we assume that the electric charge $q$  carried by the classical test particle vanishes together with rest mass $m$ and we thus end up with a null geodesic world line. The treatment of null geodesics is therefore the same as in the neutral case in Section II. 

The electromagnetic and gravitational perturbations of Melvin's magnetic universe have been thoroughly studied in connection with its stability~\cite{Melvin:1965zza, Thorne:1965}. Moreover, the propagation of test Dirac fields and self-interacting scalar fields on Melvin's spacetime have also received attention, see Refs.~\cite{Santos:2015esa, Brihaye:2021jop} and the references cited therein. We should also mention here many generalizations of the Bonnor-Melvin solution involving nonlinear superpositions of exact solutions of general relativity such as in Refs.~\cite{Ernst, Panov:1979di, GarMel1, GarMel2, Davidson:1999fa, Gibbons:2001sx, Ortag, Havrdova:2006gi, Kadlecova:2010je,Garfinkle:2011mp, Halilsoy:2012ea, Kastor:2015wda, Kastor:2020wsm} and the references therein. These gravitational fields include black holes, gyratons and nonlinear gravitational waves  on the Melvin spacetime background. 

To complement our description of particle motion in the Melvin magnetic universe, we discuss the propagation of charged scalar waves on the background Melvin spacetime in the following section.

\section{Charged Scalar Field}

We now turn to the study of the linear propagation of charged scalar field in the Melvin magnetic universe. Let $\epsilon$ be the amplitude of the matter waves under consideration here. The influence of this perturbation on the background geometry turns out to be of order $\epsilon^2$  and can be neglected. Thus we can consider the scalar field as a test field. For the purposes of comparison, we first start with a neutral massive scalar field. 

\subsection{Massive Scalar Field}

Consider a scalar field $\Phi$ with inertial mass $m$ in a spacetime with metric $g_{\mu \nu}$. The Lagrangian density for this matter wave can be written as 
\begin{equation}\label{S1}
\mathcal{L}_0 = -\frac{1}{2} \theta \left(g^{\mu \nu} \Phi_{,\mu} \Phi_{,\nu}  + \frac{1}{\ell^2}\,\Phi^2\right)\,,
\end{equation}
where $\theta := [-\det(g_{\alpha \beta})]^{1/2}$ and $\ell := \hbar /(mc)$ is the Compton wavelength of the particle. From the variational principle of stationary action, $\delta \int \mathcal{L}_0\,d^4x = 0$, we find the scalar wave equation
\begin{equation}\label{S2}
g^{\mu \nu} \Phi_{; \mu \nu} - \frac{1}{\ell^2}\Phi = 0\,,
\end{equation}
which can also be written as 
\begin{equation}\label{S3}
\frac{1}{\theta}\,\frac{\partial}{\partial x^\mu} \left(\theta \,g^{\mu \nu}\frac{\partial \Phi}{\partial x^\nu} \right)  - \frac{1}{\ell^2}\Phi = 0\,.
\end{equation}
For the Melvin universe, $\theta = a^4 \rho (1+\rho^2)^2$ and the wave equation reduces to 
\begin{equation}\label{S4}
 -\frac{\partial^2 \Phi}{\partial t^2}  +\frac{1}{\rho}\frac{\partial}{\partial \rho}\left(\rho \frac{\partial \Phi}{\partial \rho}\right) 
+\frac{(1+\rho^2)^4}{\rho^2}\frac{\partial^2 \Phi}{\partial \phi^2} +\frac{\partial^2 \Phi}{\partial z^2} - \frac{a^2}{\ell^2}(1+\rho^2)^2 \Phi = 0\,.
\end{equation}
The spacetime metric is invariant under translations in $t$, $\phi$ and $z$; hence, we can assume a solution of the form
\begin{equation}\label{S5}
 \Phi = e^{-i \omega t + i \mu \phi +i k_z z} f(\rho)\,,
\end{equation}
where $\omega$, $\mu$ and $k_z$ are constants. The wave function $\Phi$ representing a neutral scalar field is real and satisfies a linear wave equation; for  the sake of simplicity, $\Phi$ can be written in  complex form~\eqref{S5} with the convention that its real part has physical significance.  The background static configuration is invariant under $\phi \mapsto \phi + 2 \pi$; therefore, we conclude that $\mu = n = 0, \pm 1, \pm 2, \pm 3,\cdots$. Then, Eq.~\eqref{S4} implies
\begin{equation}\label{S6}
\frac{1}{\rho}\frac{d}{d \rho}\left(\rho \frac{d f}{d \rho}\right) - U(\rho) f + (\omega^2-k_z^2) f = 0\,, \quad U = \frac{a^2}{\ell^2}(1+\rho^2)^2+ \frac{n^2}{\rho^2}(1+\rho^2)^4\,.
\end{equation}
This ordinary differential equation of second order has a regular singularity at $\rho = 0$ and an irregular singularity at $\rho = \infty$; in this respect, Eq.~\eqref{S6} is similar to Bessel's equation that is the solution to Laplace's equation in cylindrical coordinates. 

Let us write the radial function $f(\rho)$ as
\begin{equation}\label{S7}
f := \frac{S(\rho)}{\rho^{1/2}}\,;
\end{equation}
then, we find a radial Schr\"odinger-like equation given by
\begin{equation}\label{S8}
\frac{d^2S}{d\rho^2} + [(\omega^2-k_z^2) - W(\rho)] S= 0\,, \quad W = U -\frac{1}{4\rho^2}\,.
\end{equation} 
It is interesting to note that up to a constant factor, $U(\rho)$ is essentially the same as $V(\rho)$ that we found in Eq.~\eqref{T3} for the geodesic motion of neutral test particles in the Melvin universe; indeed, $V(\rho) = \ell^2 U(\rho)$ if we identify $P_\phi$ with $\ell\, n = \hbar \,n/(mc)$.  This potential, except for $n = 0$,  is qualitatively similar to a harmonic oscillator potential with an infinite number of bound states corresponding to stable orbits of massive particles. While the classical motion of particles is in effect confined to the regime between the turning points, evanescent scalar waves exist in the regions where particle motion is forbidden.  These qualitative considerations apply even in the case of  a massless field when $\ell \to \infty$. 

Let us now consider the special case of $n = 0$, in which the dependence on the azimuthal angle disappears and waves can propagate radially as well as parallel to the symmetry axis. It is useful to define 
$\omega_0 > 0$ in Eq.~\eqref{S6} such that 
\begin{equation}\label{S9}
\omega_0^2 := \omega^2-k_z^2-\frac{a^2}{\ell^2}\,;
\end{equation} 
then, Eq.~\eqref{S6} for $n=0$ can be written as
\begin{equation}\label{S10}
\frac{1}{\rho}\frac{d}{d \rho}\left(\rho \frac{d f_0}{d \rho}\right) -  \frac{a^2}{\ell^2} \rho^2(2+\rho^2) f_0 + \omega_0^2 f_0 = 0\,.
\end{equation}
In the massless limit, $\ell = \infty$, Eq.~\eqref{S10} for $f_0$ has a solution that is regular on the symmetry axis and is given up to a multiplicative constant by a Bessel function of order zero, namely,  
$J_0(\omega_0 \rho)$. The massless scalar wave amplitude $f_0$ oscillates along the radial direction away from the symmetry axis $(\rho = 0)$ with decreasing amplitude and vanishes as $(\omega_0 \rho)^{-1/2}$ when $\rho \to \infty$. In the massive case, $\ell \ne \infty$, the behavior of the scalar wave amplitude $f_0$ is the same as in the massless case very near the symmetry axis ($\rho \to 0$), but is drastically different far from the symmetry axis. Indeed, $f_0(\rho)$ diverges exponentially as $\rho\to \infty$. There does not seem to be any connection between this result and the $P_\phi = 0$ case of particle motion. 

The correspondence principle establishes a connection between wave mechanics and particle mechanics in the eikonal limit of certain large numbers such as $|n| \gg 1$ in our case.  Therefore, in the case of $n = 0$, we should not expect any simple connection between wave propagation and particle motion.

\subsection{Charged Matter Waves} 

Consider now a scalar field that represents the matter wave corresponding to a particle of mass  $m$ and electric charge $q$. In the presence of an electromagnetic field, the minimal coupling of the particle with the electromagnetic field is effected via the replacement of the momentum of the particle $\mathcal{P}_\mu$  by $\mathcal{P}_\mu -(q/c) A_\mu$, where $A_\mu$ is the electromagnetic vector potential. This prescription is equivalent to replacing $\partial_\mu$ by $\partial_\mu - i (q/\hbar c) A_\mu$, since $\mathcal{P}_\mu = -i \hbar \,\partial_\mu$. The Lagrangian density for the complex scalar field corresponding to the charged particle can be obtained from the generalization of Eq.~\eqref{S1} and is given by~\cite{HE} 
\begin{equation}\label{S12}
\mathcal{L}_q = -\frac{1}{2}\theta \left[ g^{\mu \nu} (\Psi_{,\mu} - i \tfrac{q}{\hbar c} A_\mu \Psi) ( \bar{\Psi}_{,\nu} + i \tfrac{q}{\hbar c} A_\nu \bar{ \Psi}) + \frac{m^2c^2}{\hbar^2}\,\Psi\bar{\Psi}\right]\,,
\end{equation}
where $\bar{\Psi}$ is the complex conjugate of $\Psi$. From $\delta \int \mathcal{L}_g\,d^4x = 0$, we find via independent variations of $\bar{ \Psi}$ and $\Psi$, 
\begin{equation}\label{S13}
g^{\mu \nu} \Psi_{; \mu \nu} - i \tilde{q} A^{\alpha} ( 2 \Psi_{,\alpha} - i \tilde{q} A_\alpha \Psi) -i \tilde{q} A^{\alpha}{}_{;\alpha} \Psi = \frac{1}{\ell^2}\Psi\,
\end{equation}
and its complex conjugate, respectively. Here, $\tilde{q} := q/(\hbar c)$. The Lagrangian density~\eqref{S12} is real by construction; furthermore, in the absence of the electromagnetic field, Eq.~\eqref{S13} reduces to the scalar wave equation, as expected. 

Let us briefly indicate here, via the quasiclassical approximation, the general connection between the wave equation~\eqref{S13} and the motion of classical charged particles in spacetime. For the solution of Eq.~\eqref{S13} in this case, we assume an eikonal (``WKB") expansion of the form
\begin{equation}\label{Sa}
 \Psi = e^{i \,\mathbb{S}(x)/\hbar} \sum_{j = 0}^{\infty} \hbar^j \mathcal{F}_j(x)\,,
\end{equation}
where the action $\mathbb{S}(x)$ is a \emph{real} scalar function of the spacetime coordinates and $\hbar$ formally approaches zero. In the asymptotic series in Eq.~\eqref{Sa}, $\mathcal{F}_j$, $j = 0, 1, 2, ...$,  are slowly varying scalar amplitudes such that  $\mathcal{F}_0 \ne 0$ by assumption. The substitution of ansatz~\eqref{Sa} in the wave equation~\eqref{S13} results in an expansion in increasing powers of $\hbar$ beginning with $\hbar^{-2}$. The equation of particle motion is obtained by setting the coefficient of $\hbar^{-2}$ term equal to zero. Similarly, the other equations describe the evolution of $\mathcal{F}_j(x)$, $j = 0, 1, 2, ...$, along the particle trajectory. In the present case, the vanishing of the coefficient of $\hbar^{-2}$ term results in
\begin{equation}\label{Sb}
 g^{\mu \nu} \mathbb{S}_{,\mu} \mathbb{S}_{,\nu} -2q A^\alpha \mathbb{S}_{,\alpha} + q^2 A^\alpha A_\alpha = - m^2\,,
\end{equation} 
since $\mathcal{F}_0$ does not vanish by assumption. Equation~\eqref{Sb} can be written as
\begin{equation}\label{Sc}
 g^{\mu \nu} (\mathbb{S}_{,\mu} - q A_\mu) (\mathbb{S}_{,\nu} - q A_\nu)  = - m^2\,,
\end{equation} 
which is the Hamilton-Jacobi equation for the motion of  a test particle of charge $q$ and mass $m$. Indeed, let 
\begin{equation}\label{Sd}
 \mathfrak{P}_\mu = \frac{\partial \, \mathbb{S}}{\partial x^\mu}\,,
\end{equation}
be the canonical momentum of the particle; then, 
\begin{equation}\label{Se}
 \mathfrak{P}_\mu = mU_\mu + q A_\mu\,,
\end{equation} 
where $mU_\mu$ is the kinetic momentum and $U^\mu = dx^\mu/d\tau$ is the 4-velocity of the charged test particle. Taking the covariant derivative of Eq.~\eqref{Sc}, we find
\begin{equation}\label{Sf}
 g^{\mu \nu} (\mathbb{S}_{,\mu} - q A_\mu)_{; \alpha} (\mathbb{S}_{,\nu} - q A_\nu)  = 0\,,
\end{equation} 
or
\begin{equation}\label{Sg}
 (\mathbb{S}_{;\mu \alpha} - q A_{\mu; \alpha})U^\mu  = 0\,.
\end{equation} 
The Levi-Civita connection is symmetric; hence, $\mathbb{S}_{;\mu \alpha} = \mathbb{S}_{;\alpha \mu}$. Equation~\eqref{Sg} can then be expressed as 
\begin{equation}\label{Sh}
 (\mathfrak{P}_{\alpha})_{; \mu} U^\mu = q A_{\mu; \alpha}U^\mu\,,
\end{equation}
or, using Eq.~\eqref{Se}, 
\begin{equation}\label{Si}
 (m U_\alpha +q A_\alpha)_{; \mu} U^\mu = q A_{\mu; \alpha}U^\mu\,,
\end{equation}
which finally results in the Lorentz force law~\eqref{M1}, namely,
\begin{equation}\label{Si}
 m \frac{D U_\alpha}{d\tau} = q F_{\alpha \mu}U^\mu\,.
\end{equation}  

For the Melvin magnetic universe, wave equation~\eqref{S13} reduces to 
\begin{align}\label{S14}
 {}&-\frac{\partial^2 \Psi}{\partial t^2}  +\frac{1}{\rho}\frac{\partial}{\partial \rho}\left(\rho \frac{\partial \Psi}{\partial \rho}\right) 
+\frac{(1+\rho^2)^4}{\rho^2}\frac{\partial^2 \Psi}{\partial \phi^2} + 2i a \tilde{q}\,\frac{(1+\rho^2)^3}{\rho^2}\frac{\partial \Psi}{\partial \phi}+\frac{\partial^2 \Phi}{\partial z^2} \nonumber  \\
{}& - \frac{a^2}{\ell^2}(1+\rho^2)^2\left(1+\frac{\ell^2\tilde{q}^2}{\rho^2}\right) \Psi = 0\,.
\end{align}
Next, with an ansatz of the form of Eq.~\eqref{S5}, namely,
\begin{equation}\label{S15}
 \Psi = e^{-i \omega t + i n \phi +i k_z z}\, \psi(\rho)\,,
\end{equation} 
we find, as before, 
\begin{equation}\label{S16}
\frac{1}{\rho}\frac{d}{d \rho}\left(\rho \frac{d \psi}{d \rho}\right) - \mathcal{U}(\rho) \psi + (\omega^2-k_z^2) \psi = 0\,, 
\end{equation}
where
\begin{equation}\label{S17}
\mathcal{U} = \frac{a^2}{\ell^2}(1+\rho^2)^2 + \frac{(1+\rho^2)^2}{\rho^2}[(1+\rho^2)n + a \tilde{q}]^2\,.
\end{equation}
Just as in the neutral case, $\rho = 0$ is a regular singular point and the point at infinity is an essential singularity of the ordinary differential equation for $\psi(\rho)$.  With 
\begin{equation}\label{S18}
\psi(\rho) := \frac{\mathcal{Q}(\rho)}{\rho^{1/2}}\,
\end{equation}
 we find, as before,  a radial Schr\"odinger-like equation for $\mathcal{Q}$ given by
\begin{equation}\label{S19}
\frac{d^2\mathcal{Q}}{d\rho^2} + [(\omega^2-k_z^2) - \mathcal{W}(\rho)] \mathcal{Q}= 0\,, \quad \mathcal{W} = \mathcal{U} -\frac{1}{4\rho^2}\,.
\end{equation} 
At the boundaries of the range of $\rho: 0 \to \infty$, the dominant terms in the potential $\mathcal{W}(\rho)$ are given by 
\beq
\label{S20}
\mathcal{W}(\rho)\bigg|_{\rho\to 0 }\sim \left[(n+a\tilde q)^2-\frac{1}{4}\right] \frac{1}{\rho^2}\,, \qquad  \mathcal{W}(\rho)\bigg|_{\rho\to \infty } \sim  n^2 \rho^6 + O(\rho^4)\,;
\eeq
therefore, the general behavior of the potential for $2|n + a \tilde{q}| > 1$ appears to be roughly similar to a harmonic oscillator potential with a single minimum; however, the graph of  
$\mathcal{W}(\rho)$ reveals that the bottom of the potential near the minimum can be of considerable width. To show that there is indeed only one minimum for $2|n + a \tilde{q}| > 1$, we note that  $d\mathcal{W}/d\rho = 0$ can be written as 
\beq
\label{S21}
\left[(n+a\tilde q)^2-\frac{1}{4}\right] \frac{1}{\rho^4} + n^2 (1+\rho^2) (3 + \rho^2) =  \frac{a^2}{\ell^2}(1+\rho^2) + (3n + a\tilde q + 2n \rho^2)^2\,.
\eeq
Inspection of the graphs for both sides of this relation makes it clear that they can  cross only once. On the other hand, for $2|n + a \tilde{q}| \le 1$, the two sides of Eq.~\eqref{S21} do not cross; therefore, the potential does not have an extremum; in fact,  $\mathcal{W}(\rho)$ is a monotonically increasing function of $\rho^2$. The nature of $\mathcal{W}(\rho)$ is illustrated in Figure 3. 

\begin{figure}
\includegraphics[scale=0.4]{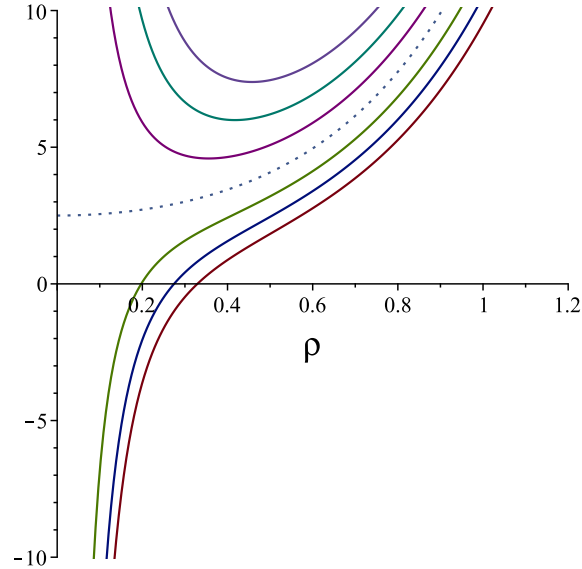}
\caption{\label{fig_potentials}  Plot of the potential $\mathcal{W}(\rho)$ versus $\rho$ for $a = \ell$, $n = 1$ and $a\tilde{q}= (-5+N)/10$, where $N$ is an integer that ranges from $-3$ (lowest curve) to $3$ (highest curve). The dotted curve is the separatrix and  corresponds to the critical value $N=0$.
A curve with a minimum belongs to the case $(n+a\tilde q)^2>1/4$, while a curve with a flex belongs to the case $(n+a\tilde q)^2<1/4$. These are separated by the separatrix, which is associated with the critical case $(n+a\tilde q)^2=1/4$. 
}
\end{figure}

The correspondence with charged particle motion can be established for $|n| \gg 1$, which is compatible with  $2|n + a \tilde{q}| > 1$; then, the Schr\"odinger equation~\eqref{S19} has bound states corresponding to the stable helical motions of charged particles with elliptic or circular paths in the $(x, y)$ plane. However, for $2|n + a \tilde{q}| \le 1$, no clear connection with charged particle motion exists. 
Finally, in the special case $n + a \tilde{q} = 0$, the solution of Eq.~\eqref{S16}, i.e. $\psi$, very close to the axis of symmetry behaves like $J_0(\omega_0 \rho)$, while  it diverges exponentially for $\rho \to \infty$.

The helical motion of neutral and charged particles up and down the long axis of cylindrical symmetry of Melvin's magnetic universe is certainly reminiscent of the highly collimated double-jet configuration of the extended synchrotron emitting radio sources associated with active galactic nuclei~\cite{Punsly}. However, the Melvin magnetic universe is static and the energy of a  particle as measured by fiducial observers is constant. In relativistic jets encountered in astrophysics, particles can be accelerated to almost the speed of light. This circumstance provides the motivation to explore the physics of a dynamic Melvin universe in which the characteristic length scale of the universe becomes dependent upon time; that is, $a \to \mathbb{A}(t)$. Furthermore,  the stability of Melvin's magnetic universe is due to the fact that magnetic tension can balance gravitational attraction. In a recent paper~\cite{Tsagas:2020lal}, the implications of this circumstance for gravitational collapse have been studied. Indeed, to induce gravitational collapse in the Melvin universe, it appears that it is in general necessary to violate the weak energy condition; however, this can be avoided in certain circumstances. These issues will be discussed in the next section.

\section{A Simple Conformal Collapse Scenario}

The purpose of this section is to explore the possibility that Melvin's universe could undergo gravitational collapse. One might try making $B_0$ in metric~\eqref{I1} time dependent. Another possibility is to consider the same form as metric~\eqref{I4}, but assume that the parameter $a$ depends upon time. Of all the possibilities that we have tried, the simplest is to let $a \to \mathbb{A}(t)$,
\beq
\label{G1}
ds^2=\mathbb{A}^2(t) \left[(1+\rho^2)^2(-dt^2+d\rho^2+dz^2)+\frac{\rho^2}{(1+\rho^2)^2}d\phi^2 \right]\,,
\eeq
where $\mathbb{A}$ is a \emph{linear} function of time $t$. That is, 
\beq
\label{G2}
\mathbb{A}(t) = a - b\, t\,,
\eeq
where $b > 0$ is a constant length. The Melvin magnetic universe is static and has an infinite lifetime; however, the dynamic Melvin universe introduced here has a finite lifetime starting from $t= 0$, when the universe begins to collapse gravitationally until this process is completed when $\mathbb{A} = 0$ in Eq.~\eqref{G2} and the universe disappears at
\beq   \label{G3}
t=t_*=\frac{a}{b}\,.
\eeq

Let $e_{\hat \alpha}$ be the natural orthonormal frame~\eqref{I15}  adapted to the reference observers spatially at rest with $a \to \mathbb{A}(t)$. In this case,  instead of Eq.~\eqref{I18}, we find the measured components of the acceleration tensor in the \emph{dynamic} Melvin universe are given by
\beq    \label{G4}
\mathbf{g}_D=\frac{2\rho}{\mathbb{A}(1+\rho^2)^2}\,e_{\hat1}\,,\qquad {\boldsymbol \omega}_D=0\,.
\eeq
Therefore, as $t \to t_*=a/b$, the radial attraction of gravity diverges; that is, all test particles are attracted toward the axis of symmetry. In particular, the volume expansion~\cite{HE} of the congruence of fiducial observers is given by
\beq    \label{G4a}
\Theta =-\frac{3 b}{{\mathbb A}^2(1+\rho^2)}\,.
\eeq

The Einstein tensor corresponding to metric~\eqref{G1} can be employed to find the source of the dynamic Melvin universe. The result is the new traceless energy-momentum tensor $\mathbb{T}^{\mu}{}_{\nu}$, which has diagonal elements 
\beq    \label{G5}
8\pi \mathbb{T}^{\mu}{}_{\nu} =\frac{4}{{\mathbb A}^2(1+\rho^2)^4}{\rm diag}[-1,1,1,-1] +\frac{b^2}{{\mathbb A}^4(1+\rho^2)^2}{\rm diag}[-3,1,1,1]\,
\eeq
and nonzero off-diagonal elements
\beq      \label{G6}
8\pi \mathbb{T}^t{}_\rho =-8\pi \mathbb{T}^\rho{}_t =\frac{4b \rho}{{\mathbb A}^3(1+\rho^2)^3}\,.
\eeq
In general, this energy-momentum tensor violates the weak energy condition. Let $\chi := \mathbb{T}_{\mu \nu} W^\mu W^\nu$ for an arbitrary timelike (or null) vector $W^\alpha$  such that $W^\mu W_\mu \le 0$; then, the \emph{weak energy condition} is satisfied provided $\chi \ge 0$. We note that  $\mathbb{T}_{\mu \nu}$ is traceless; therefore, the weak and strong energy conditions coincide in this case. For the case of the dynamic Melvin universe, $\chi \ge 0$ along the symmetry axis or when $W^\mu$ represents the 4-velocity vector of the fiducial observers. However, the weak energy condition is in general violated; for instance, $\chi < 0$ for an outgoing null vector in the radial direction when $\mathbb{A}/b = 2$ and $\rho = 3$.  On the other hand, it is possible to find special situations where these energy conditions are indeed satisfied; for instance, we show in Appendix C that $\chi \ge 0$ for a dynamic Melvin universe with $t_* \le \sqrt{21}/4$;  that is, the conformal collapse scenario could have a satisfactory source if the age of the resulting dynamic Melvin universe is sufficiently short.  

We will forgo the attempt to provide a complete physical characterization of our gravitational source $\mathbb{T}_{\mu \nu}$, except to note that circumstances exist, as shown in Appendix C, where the source satisfies the weak and strong energy conditions. In Section VI, we will explore the possibility that $\mathbb{T}_{\mu \nu}$ could be partly due to electromagnetic fields left over from the Melvin magnetic universe. 

The essential singularity of this gravitational field occurs at the absolute end of the collapsing universe when $\mathbb{A} = 0$ at $t_* = a/b$. As expected, this singularity shows up  in the Kretschmann and Weyl invariants, namely, 
\beq   \label{Ga}
R^{\alpha\beta\mu\nu}R_{\alpha\beta\mu\nu} =\frac{1}{{\mathbb A}^4 (1+\rho^2)^4}\left[\frac{64 (3\rho^4-6\rho^2+5)}{(1+\rho^2)^4} + \frac{24 b^4}{{\mathbb A}^4 }-\frac{64 b^2 (\rho^2-1)}{{\mathbb A}^2 (1+\rho^2)^2}\right]\,
\eeq
and
\beq  \label{Gb}
 C_{\alpha\beta\gamma\delta}C^{\alpha\beta\gamma\delta}=\frac{192(\rho^2-1)^2}{\mathbb{A}^4(1+\rho^2)^8}\,,
\eeq
respectively. 

Let us next  study the motion of future-directed timelike paths of free test particles in this dynamic spacetime and explore the possibility of jet formation. It is important to point out that double-jet structures can  occur even without the presence of electromagnetic fields; indeed, in Refs.~\cite{Chicone:2010aa, Chicone:2010xr, Chicone:2011ie, Bini:2014esa, Bini:2017uax, Bini:2017qnd} cosmic double-jet patterns are  exhibited and studied in certain dynamic spacetimes.  

\subsection{Geodesic Motion}

The metrics of the dynamic and static Melvin universes are connected via the conformal factor     
\beq  \label {U1}  
\Omega  = \frac{\mathbb{A}}{a} = 1 - \frac{t}{t_*}\,;
\eeq
that is, $\tilde{g}_{\mu \nu} = \Omega^2 g_{\mu \nu}$.  Imagine any future-directed timelike path in the dynamic Melvin universe. The proper time $\tau$ along the path is given by
\beq  \label {U2}  
d\tau^2 = - \tilde{g}_{\mu \nu}\, dx^\mu dx^\nu\,,
\eeq
which tends to zero as $t \to t_*$. Therefore, as the singularity is approached, all future-directed timelike paths approach null paths. We explore this circumstance here via geodesics and note that null geodesics are conformally invariant, which means that as $t \to t_*$, future-directed timelike geodesics of the dynamic Melvin universe turn into null geodesics of the static Melvin universe.   

As is well known, the geodesic equations of motion of a test particle can be obtained from the Lagrangian
\beq   \label{U3}
L = -\frac{1}{2} \tilde{g}_{\mu \nu} \frac{dx^\mu}{d\tau}\frac{dx^\nu}{d\tau}\,.
\eeq
Let us consider the motion of a neutral test particle with 4-velocity  $u^\mu = dx^\mu/d\tau$, where $x^\mu = (t, \rho, \phi, z)$.  
The existence of $\phi$ and $z$ Killing vectors implies 
\beq    \label{U4}
\mathbb{A}^2(t)\rho^2 (1+ \rho^2 )^{-2} \frac{d\phi}{d\tau} = \mathbb{P}_\phi\,, \quad \mathbb{A}^2(t) (1+ \rho^2 )^2 \frac{dz}{d\tau} = \mathbb{P}_z\,, \quad \frac{d\phi}{d z}= \frac{\mathbb{P}_\phi}{\mathbb{P}_z}\frac{(1+ \rho^2 )^4}{\rho^2}\,,
\eeq
where  $\mathbb{P}_\phi$ and $\mathbb{P}_z \ne 0$ are constants of the motion; indeed, they are the particle's canonical momenta (per unit mass) along the azimuthal and vertical directions, respectively. Moreover, $u^\mu u_\mu = -1$ implies
\beq     \label{U5}
\left(\frac{dt}{d\tau}\right)^2 - \left(\frac{d\rho}{d\tau}\right)^2 - \left(\frac{dz}{d\tau}\right)^2 - \frac{\rho^2}{(1+\rho^2)^4}\left(\frac{d\phi}{d\tau}\right)^2= \frac{1}{\mathbb{A}^2(t) (1+ \rho^2 )^2}\,.
\eeq

We are interested in free test particles moving up and down parallel to the symmetry axis; therefore we assume $\mathbb{P}_z \ne 0$. The Euler-Lagrange equations of motion corresponding to $t$ and $\rho$ can be expressed as
\beq     \label{U6}
\frac{d}{d\tau} \left[\mathbb{A}^2(t) (1+ \rho^2 )^2 \frac{dt}{d\tau}\right]  = - \frac{b}{\mathbb{A}}\,
\eeq
and
\beq       \label{U7}
\frac{d}{d\tau} \left[\mathbb{A}^2(t) (1+ \rho^2 )^2 \frac{d\rho}{d\tau}\right]  = -2\, \frac{\rho}{1+\rho^2} + \frac{\mathbb{P}_\phi^2}{\mathbb{A}^2(t)} \frac{(1+\rho^2)(1-3\rho^2)}{\rho^3}\,,
\eeq 
respectively. Here,  we have employed  Eqs.~\eqref{U4} and~\eqref{U5}.  Using $\mathbb{P}_z \ne 0$, we can simply express  the Euler-Lagrange  equations of motion as 
\beq     \label{U8}
\frac{d^2t}{dz^2} = - \frac{b}{\mathbb{P}_z^2}\, \mathbb{A} (1+\rho^2)^2\,
\eeq
and 
\beq     \label{U9}
\frac{d^2\rho}{dz^2} = - \frac{2}{\mathbb{P}_z^2} \,\mathbb{A}^2 \rho (1+\rho^2) + \frac{\mathbb{P}_\phi^2}{\mathbb{P}_z^2} \frac{(1+\rho^2)^3(1-3\rho^2)}{\rho^3}\,,
\eeq
respectively. To these equations, we must add 
\beq     \label{U10}
\left(\frac{dt}{dz}\right)^2 - \left(\frac{d\rho}{dz}\right)^2 = 1 +  \frac{\mathbb{A}^2(t) (1+ \rho^2 )^2}{\mathbb{P}_z^2} +  \frac{\mathbb{P}_\phi^2 }{\mathbb{P}_z^2}\frac{(1+ \rho^2 )^4}{\rho^2}\,,
\eeq
which is obtained from Eqs.~\eqref{U4} and~\eqref{U5}. These equations of motion are symmetric with respect to $z$. 
Differentiating Eq.~\eqref{U10} with respect to $z$ and using Eqs.~\eqref{U8} and~\eqref{U9}, we find that  these equations are indeed compatible. 

Projecting the jet motion onto the tetrad of the fiducial observer, we get
\beq     \label{U11}
u_\mu e^{\mu}{}_{\hat \alpha} = u_{\hat \alpha}\,, \qquad u^{\hat \alpha} = \Gamma (1, V^{\hat 1}, V^{\hat 2}, V^{\hat 3})\,,
\eeq
where $\Gamma$ and $V^{\hat i}$, $i = 1, 2, 3$, are the Lorentz factor and components of the velocity of the particle as measured by the fiducial observer. In practice, it is sufficient to compute $\Gamma$, which in the present case is given by
\beq       \label{U12}
\Gamma = \frac{\mathbb{P}_z}{\mathbb{A}(t) (1+ \rho^2 )} \,\frac{dt}{dz}\,.
\eeq
Numerical experiments demonstrate that as $t \to t_*$,  jets of free particles experiencing helical motions up and down the symmetry axis are strongly attracted toward the symmetry axis and their Lorentz factors tend to infinity. This is illustrated in Figure 4.

\begin{figure}[h]
\includegraphics[scale=0.40]{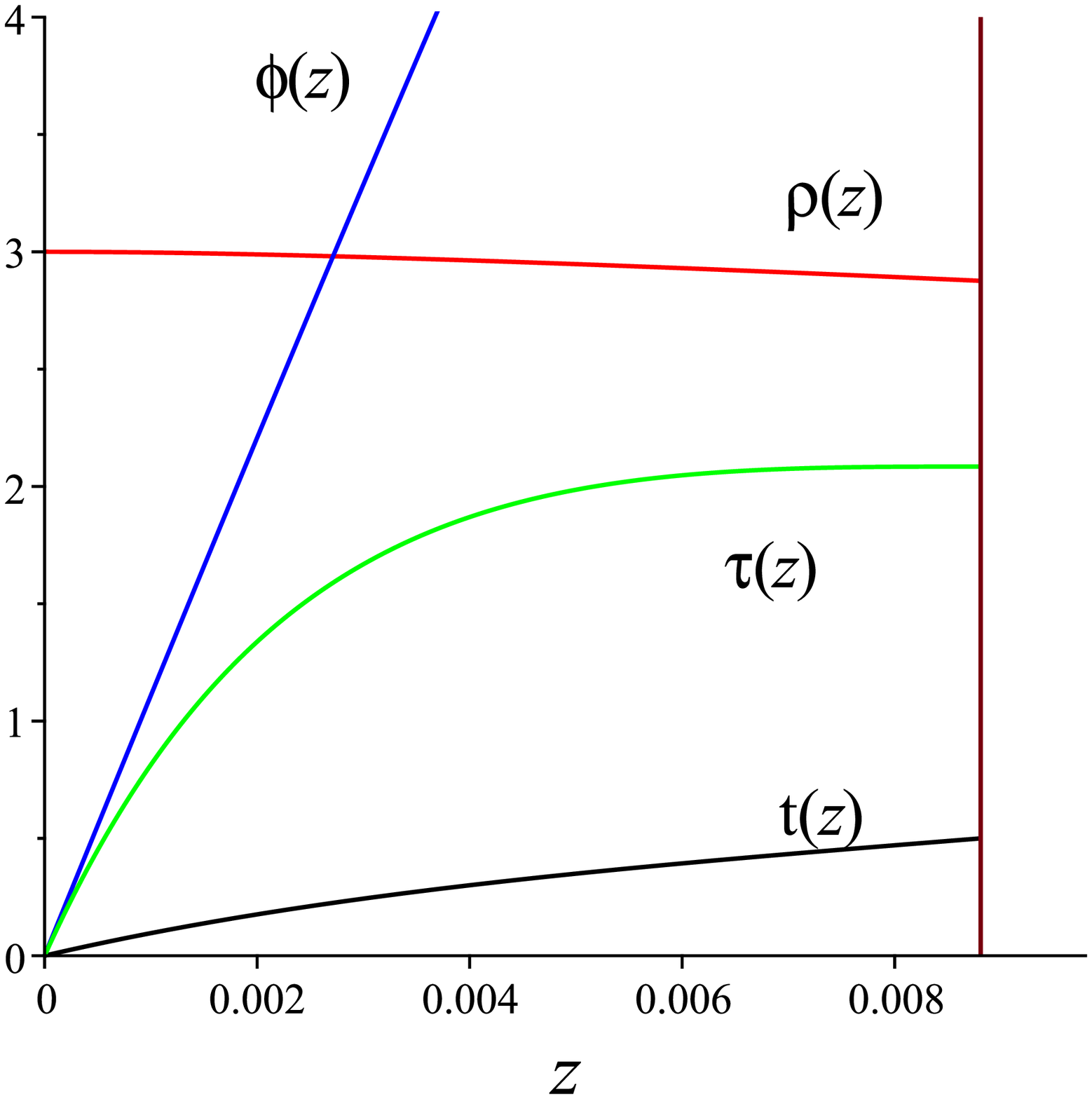}
\includegraphics[scale=0.40]{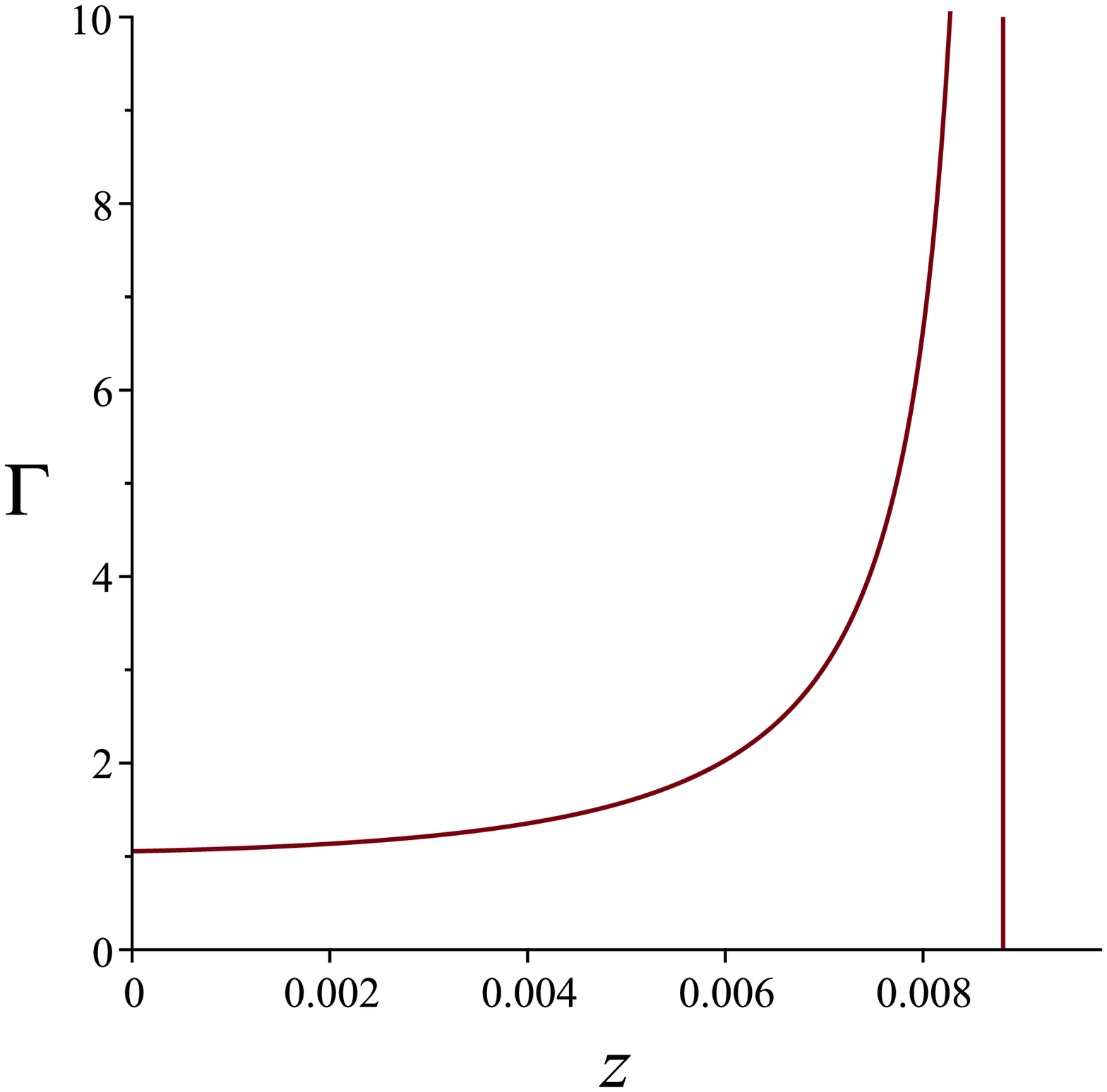}
\caption{\label{fig_t_z} Left Panel: Numerical integration of $t(z)$, $\tau(z)$, $\rho(z)$ and $\phi(z)$ using the geodesic equations of motion for a free neutral test particle. We employ parameters:
$a=1$, $b=2$, $\mathbb{P}_\phi=0.1$ and $\mathbb{P}_z=0.1$, and initial conditions:
$t(0)=0$, $\tau(0)=0$, $\rho(0)=3$, $\phi(0)=0$, $(d\rho/dz)(0)=0$ and  $(dt/dz)(0)\approx105.414$ [obtained from Eq.~\eqref{U10}].
Right Panel: Corresponding numerical integration of Lorentz factor $\Gamma(z)$. The universe ends at $t_* = 0.5$, $\tau(t_*) \approx 2.085$,
$\rho(t_*) \approx 2.876$, $\phi(t_*)  \approx  9.008$ and $z(t_*) \approx 0.009$. Units are not specified here, since the plots are purely for illustration. }
\end{figure}

\section{Electrodynamics of Conformal Collapse}

In the conformal collapse scenario, the static Melvin magnetic universe undergoes gravitational collapse at $t = 0$.  The Melvin spacetime is characterized by its magnetic source; therefore, a dynamic Melvin universe should naturally have a corresponding electromagnetic field that in the static limit would correspond to the magnetic field of Melvin's static universe. Magnetic fields are also required in connection with astrophysical jets. Naturally, the corresponding electromagnetic energy-momentum tensor $T_{\mu \nu}$ would be part of the source $\mathbb{T}_{\mu \nu}$ of the resulting dynamic Melvin universe. 
These considerations lead to two simple possible choices for the electromagnetic vector potential
\beq   \label{H1}
A=-\frac{\mathbb{A}(t)}{(1+\rho^2)}d\phi\, \qquad  {\rm or} \qquad A=-\frac{a}{(1+\rho^2)}d\phi\,.
\eeq
These coincide at $t =0$ with the vector potential of the static Melvin magnetic universe. The first possibility involves an electric field as well and will be investigated in this section, while the second possibility, suggested by the conformal invariance of Maxwell's equations,  has no associated electric field and will be studied in the next section. 

In the first possibility in Eq.~\eqref{H1}, the linear dependence of $\mathbb{A}$ upon coordinate time $t$ implies that Maxwell's equations are satisfied in this case with the electromagnetic field
\beq   \label{H2}
F = dA= \frac{b}{1+\rho^2} dt\wedge d\phi + \frac{2\rho \mathbb{A}(t)}{(1+\rho^2)^2} d\rho \wedge d\phi\,.
\eeq
Therefore, in addition to the time-dependent magnetic field component parallel to the symmetry axis, we also have an azimuthal electric field component as a consequence of Faraday's law.  Indeed, the measured magnitudes of these fields are
\beq   \label{H3}
E_b = \frac{b}{\mathbb{A}^2 \rho(1+\rho^2)}\,,  \qquad    B_b = \frac{2}{\mathbb{A}(t)(1+\rho^2)}\,
\eeq
and the new algebraic electromagnetic invariants $I_1$ and $I_2$ are given by $B_b^2-E_b^2$ and $0$, respectively. 

The measured electric field diverges as $1/\rho$ on the symmetry axis ($\rho = 0$).  This divergence is mitigated by the issue of the measurability of the electric field.  According to Bohr and Rosenfeld, classical electromagnetic fields cannot be measured at a spacetime event; instead, a spacetime averaging is required on the basis of the Lorentz force law. To see this, let us place a small extended classical charged particle in an electric field and measure the change in the momentum of the particle over a suitable period of time. The impulse can be calculated using the Lorentz force law. The result of this classical experiment is in effect the measurement of the average of the electric field over the spacetime domain that consists of the volume of the charged particle multiplied by the period of the measurement~\cite{BR}. The symmetry axis of the Melvin universe is elementary flat; therefore, the integration of the ``measured" electric field in Eq.~\eqref{H3} over the cylindrical volume element, $\rho\, d\rho\wedge d\phi\wedge dz$, that is involved in the determination of the actual measured field at the axis will be free of the field singularity.    

It may be interesting to see the additional source that is needed to cause the collapse of the Melvin universe. We can express  the gravitational source as
\beq   \label{H4}
\mathbb{T}^{\mu}{}_{\nu} = T^\mu{}_\nu+\delta T^\mu{}_\nu\,,
\eeq
where $T^{\mu}{}_ \nu$ is the traceless energy-momentum tensor  that corresponds to the electromagnetic field~\eqref{H2}. Indeed,   $T^{\mu}{}_{\nu}$ has diagonal components
\beq \label{H5}
8 \pi \, T^\alpha{}_\beta=\frac{4}{\mathbb{A}^2(1+\rho^2)^4}\,{\rm diag}[-1,1,1,-1] + \frac{b^2}{\rho^2\mathbb{A}^4(1+\rho^2)^2}\,{\rm diag}[-1,1,-1,1]\,
\eeq 
and off-diagonal components
\beq \label{H6}
8 \pi \, T^{t}{}_{\rho} = - 8 \pi \, T^{\rho}{}_{t}  = - \frac{4b}{\rho \mathbb{A}^3(1+\rho^2)^3}\,.
\eeq
The additional tracefree stress-energy tensor $\delta T^{\mu}{}_{\nu}$ that helps induce the collapse of the Melvin universe  has nonzero  components given by
\bea   \label{H7}
8\pi\delta T^{t}{}_{t}&=&-\frac{b^2(3\rho^2-1)}{\rho^2 \mathbb{A}^2(1+\rho^2)^2}\,,\nonumber\\
8\pi\delta T^{\rho}{}_\rho&=&\frac{b^2(\rho^2-1)}{\rho^2 \mathbb{A}^4(1+\rho^2)^2}=8\pi\delta T^{z}{}_z\,,\nonumber\\
8\pi\delta T^{\phi}{}_\phi&=&\frac{b^2}{\rho^2 \mathbb{A}^4(1+\rho^2)}\,,\nonumber\\
8\pi\delta T^{t}{}_\rho&=&\frac{4b}{\rho \mathbb{A}^3(1+\rho^2)^2} =  -8\pi T^{\rho}{}_t\,.
\eea
We note that $\delta T^{\mu\nu}{}_{;\nu}=0$; moreover,  as expected, its components all vanish when $b\to 0$, i.e. when the static Melvin magnetic universe solution is recovered.

\subsection{Charged Particles in Melvin Universe with $A=-\mathbb{A}(t)\,d\phi/(1+\rho^2)$}

We turn now to the motion of charged particles in the dynamic Melvin universe under consideration in this section. Using much of the same notation and conventions as in Section III, we are interested in the solution of the Lorentz force equation.  It follows from Eqs.~\eqref{M1}--\eqref{M3} that 
\beq
\label{V1}
\frac{d(K^\alpha U_\alpha)}{d\tau}=  \hat{q} F_{\mu \nu} K^\mu U^\nu\,,
\eeq
where $K^\mu$ is a Killing vector field and $U^\mu= dx^\mu /d\tau$ is the 4-velocity of the charged particle. The dynamic Melvin metric has Killing vectors $\partial_\phi$ and $\partial_z$. For axial symmetry, Eq.~\eqref{V1}  implies
\beq    \label{V2}
\frac{d}{d\tau}\left[\frac{\mathbb{A}^2\rho^2}{(1+\rho^2)^2} \frac{d\phi}{d\tau}\right] = \hat{q} F_{\phi t} \frac{dt}{d\tau} + \hat{q} F_{\phi\rho} \frac{d\rho}{d\tau}\,,
\eeq
or
\beq    \label{V3}
\frac{\mathbb{A}^2\rho^2}{(1+\rho^2)^2} \frac{d\phi}{d\tau} -\hat{q} \frac{\mathbb{A}}{1+\rho^2} = \tilde{P}_\phi\,,
\eeq
which is a simple generalization of Eq.~\eqref{M6} to the dynamic case. Therefore,
\beq    \label{V3a}
\frac{d\phi}{d\tau} =\frac{1+\rho^2}{\mathbb{A}^2\rho^2}\,[\hat{q}\mathbb{A} +(1+\rho^2)\tilde{P}_\phi]\,,
\eeq
where $\tilde{P}_\phi$ is a constant of the motion. When the term involving the electric charge is dominant in Eq.~\eqref{V3a}, a particle with positive (negative) charge tends to move in the positive (negative) sense about the symmetry axis. Similarly, for the translational symmetry along the axis of cylinder, we have 
\beq    \label{V4}
 \mathbb{A}^2 (1+\rho^2)^2 \frac{dz}{d\tau} = \tilde{P}_z\,,
\eeq
which generalizes Eq.~\eqref{M4}. Here, $ \tilde{P}_z$ is another constant of the motion. Furthermore, Eq.~\eqref{U5} holds in this case as well, since $U^\mu U_\mu = -1$; therefore, using Eqs.~\eqref{V3a} and~\eqref{V4} in Eq.~\eqref{U5} results in
\beq     \label{V5}
\left(\frac{dt}{d\tau}\right)^2 - \left(\frac{d\rho}{d\tau}\right)^2 = \frac{1}{\mathbb{A}^2(t) (1+ \rho^2 )^2} + \frac{\tilde{P}_z^2}{\mathbb{A}^4(t) (1+ \rho^2 )^4}+ \frac{[(1+\rho^2)\tilde{P}_\phi+ \hat{q} \mathbb{A}]^2}{\mathbb{A}^4(t) \rho^2(1+ \rho^2 )^2}\,.
\eeq
The components of the 4-velocity vector $U^\mu$ are fully determined via the above equations once we find $\rho(\tau)$. In this connection, we consider the radial component of the Lorentz force equation, namely, 
\beq
\label{V6}
\frac{dU^1}{d\tau}+\Gamma^1{}_{\alpha\beta}U^\alpha U^\beta=  \hat{q} F^1{}_{\nu}U^\nu\,.
\eeq
 This equation can be written as
\begin{align}
\label{V7}
\nonumber \frac{d^2\rho}{d\tau^2}{}&+\frac{2\rho}{1+\rho^2}\left[ \left(\frac{dt}{d\tau}\right)^2 + \left(\frac{d\rho}{d\tau}\right)^2- \left(\frac{dz}{d\tau}\right)^2\right] - 2\frac{b}{\mathbb{A}}\frac{dt}{d\tau}\frac{d\rho}{d\tau} \\
{}& -\rho\, \frac{1-\rho^2}{(1+\rho^2)^5} \left(\frac{d\phi}{d\tau}\right)^2 = 2 \hat{q} \frac{\rho}{\mathbb{A}(t) (1+ \rho^2 )^4}\frac{d\phi}{d\tau}\,.
\end{align}

We have studied the behavior of charged particles via numerical solutions of the above equations of motion. In conformity with previous results for uncharged particles, charged particles in general follow helical paths that spiral toward the axis of cylindrical symmetry as the universe collapses. As measured by observers that are spatially at rest, charged particles move with speeds close to the speed of light and their Lorentz factors tend to  infinity  as $t \to t_*$. On the other hand, the additional electromagnetic force that a charged particle experiences tends to speed up the collapse process. In our convention, positively (negatively) charged particles tend to spiral in the positive (negative) sense around the symmetry axis. Furthermore, the influence of the azimuthal electric field tends to increase the winding number of charged particles as they spiral around and collapse onto the axis of cylindrical symmetry.

\section{Conformal Collapse Scenario with $A=- a\,d\phi/(1+\rho^2)$}

Finally, we consider the possibility that the the dynamic Melvin universe is free of the large-scale electric field and simply inherits the static magnetic field~\eqref{I10} of the Melvin magnetic universe via conformal invariance of the electromagnetic field equations. On the other hand, the magnetic field as measured by the fiducial observers spatially at rest in the dynamic Melvin universe is dependent upon time and is given by  
\beq
\label{W1}
F_{\hat \rho \hat \phi} = \frac{2a}{\mathbb{A}^2}\frac{1}{(1+\rho^2)^2}\,.
\eeq 

As in the previous section, the Lorentz force law implies in this case that
\beq    \label{W2}
\frac{d\phi}{d\tau} =\frac{1+\rho^2}{\mathbb{A}^2\rho^2}\,[a\,\hat{q} +(1+\rho^2)\bar{P}_\phi]\,,
\eeq
\beq    \label{W3}
 \mathbb{A}^2 (1+\rho^2)^2 \frac{dz}{d\tau} = \bar{P}_z\,,
\eeq
where $\bar{P}_\phi$ and $\bar{P}_z$ are constants of the motion. Moreover, the 4-velocity of the charged particle is a unit timelike vector; hence, 
\beq     \label{W4}
\left(\frac{dt}{d\tau}\right)^2 - \left(\frac{d\rho}{d\tau}\right)^2 = \frac{1}{\mathbb{A}^2(t) (1+ \rho^2 )^2} + \frac{\bar{P}_z^2}{\mathbb{A}^4(t) (1+ \rho^2 )^4}+ \frac{[(1+\rho^2)\bar{P}_\phi+ a\,\hat{q}]^2}{\mathbb{A}^4(t) \rho^2(1+ \rho^2 )^2}\,.
\eeq
The last equation that is needed is given by the radial component of the Lorentz force law; that is,  
\begin{align}
\label{W5}
\nonumber \frac{d^2\rho}{d\tau^2}{}&+\frac{2\rho}{1+\rho^2}\left[ \left(\frac{dt}{d\tau}\right)^2 + \left(\frac{d\rho}{d\tau}\right)^2- \left(\frac{dz}{d\tau}\right)^2\right] - 2\frac{b}{\mathbb{A}}\frac{dt}{d\tau}\frac{d\rho}{d\tau} \\
{}& -\rho\, \frac{1-\rho^2}{(1+\rho^2)^5} \left(\frac{d\phi}{d\tau}\right)^2 = 2 a\,\hat{q} \frac{\rho}{\mathbb{A}^2(t) (1+ \rho^2 )^4}\frac{d\phi}{d\tau}\,.
\end{align}

We have numerically integrated these equations and studied the motion of charged particles in this dynamic Melvin universe. The situation here is different from the previous section due to the absence of an azimuthal electric field; nevertheless, the main results are qualitatively the same. That is, charged particles undergo helical motions as they collapse onto the axis of symmetry and their Lorentz factors tend to infinity as $t \to t_*$. 

\section{DISCUSSION}

The constant scale factor of the Melvin magnetic universe is a length that is inversely proportional to the magnetic field strength, i.e. $a = 2c^2/(G^{1/2} B_0)$. What if this conformal factor becomes dependent upon time and the universe collapses? That is, $a \to \mathbb{A}(t)$ and, for the sake of simplicity, we let $\mathbb{A}(t) = a - b\,t$, where $b>0$ is a constant. The new dynamic universe exists from $t = 0$ to $t_* = a/b$.  As the universe collapses, the measured magnetic field strength and the attraction of gravity toward the symmetry axis keep increasing and eventually diverge. In this process of conformal collapse, double-jet structures develop, but these cosmic jets soon disappear along with the collapsing universe.  

We relax the conformal collapse condition in Appendix D and consider, for instance, the collapse of the whole horizontal space of the Melvin magnetic universe around the axis of cylindrical symmetry. Cylindrical gravitational collapse has been discussed in general relativity by a number of authors, see~\cite{Thorne, Cocke, Apostolatos:1992qqj, DiPrisco:2009zc, Herrera:2012qq, Chak} and the references cited therein.  These investigations involved, for the most part,  collapsing cylindrical systems that were matched to exterior vacuum Einstein-Rosen spacetimes in order to study the cylindrical gravitational radiation generated via gravitational collapse. On the other hand, we have been interested in the problem of jet formation in quasars and active galactic nuclei. Our preliminary results deserve further investigation.

\appendix

\section{Fermi coordinates in Melvin's universe}

Infinitely extended cylindrically symmetric universe models are not compatible with cosmological observations. To gain insight into the nature of such gravitational fields, it is useful to establish local quasi-inertial Fermi normal coordinate systems~\cite{Synge, mash77} in the neighborhoods of reference observers. For the Melvin universe, we choose the congruence of observers that are spatially at rest and carry the natural orthonormal tetrad system~\eqref{I15}. 

Let $\bar{x}^\mu(\tau)$ be the world line of a fiducial observer. At an arbitrary event along the world line with proper time $\tau$, consider the class of all spacelike geodesics that emanate normally from this event and generate a local spacelike hypersurface. Let $x^\mu$ be an event on this hypersurface; then, to this event we assign invariantly defined Fermi coordinates $X^{\hat \mu}$,  
\begin{equation}\label{A1}
X^{\hat 0} := \tau\,, \qquad X^{\hat i} := \Sigma\, \xi^\mu(\tau)\, e_{\mu}{}^{\hat i}(\tau)\,,
\end{equation} 
where $\Sigma$ is the proper length of the unique spacelike geodesic segment that connects $x^\mu$ to $\bar{x}^\mu(\tau)$ and $\xi^\mu(\tau)$ is the unit spacelike vector tangent to this  geodesic segment at 
$\bar{x}^\mu(\tau)$.  The reference observer in the Melvin universe has translational acceleration $\mathbf{g}$ given by Eq.~\eqref{I18} and its spatial frame is Fermi-Walker transported along its world line. In this case, the spacetime metric in Fermi coordinates is given by
\begin{equation}\label{A2}
ds^2 = g_{\hat \mu \hat \nu}\,dX^{\hat \mu} dX^{\hat \nu}\,,
\end{equation}
where
\begin{equation}\label{A3}
g_{\hat 0 \hat 0} = - (1+g_{\hat i} X^{\hat i})^2  - R_{\hat 0 \hat i \hat 0 \hat j}\,X^{\hat i}\,X^{\hat j}\,,
\end{equation}
\begin{equation}\label{A4}
g_{\hat 0 \hat i} = -\frac{2}{3} \,R_{\hat 0 \hat j \hat i \hat k}\,X^{\hat j}\,X^{\hat k}\,
\end{equation}
and
\begin{equation}\label{A5}
g_{\hat i \hat j} = \delta_{\hat i \hat j} -\frac{1}{3} \,R_{\hat i \hat k \hat j \hat l}\,X^{\hat k}\,X^{\hat l}\,,
\end{equation}
where  third and higher-order terms in spatial Fermi coordinates have been neglected. Moreover, 
\begin{equation}\label{A6}
R_{\hat \alpha \hat \beta \hat \gamma \hat \delta} = R_{\mu \nu \rho \sigma}\,e^{\mu}{}_{\hat \alpha}\,e^{\nu}{}_{\hat \beta}\,e^{\rho}{}_{\hat \gamma}\,e^{\sigma}{}_{\hat \delta}\,
\end{equation}
is the projection of the Riemann curvature tensor upon the fiducial observer's tetrad frame.

The Fermi coordinate system is admissible in a sufficiently narrow cylindrical domain along the reference world line; in fact, the spatial Fermi coordinates should be sufficiently small compared to the local radius of curvature of spacetime~\cite{Chicone:2005vn}.  In general, we can express Eq.~\eqref{A6} as a $6\times 6$ matrix with indices that range over the set $\{01,02,03,23,31,12\}$. In an arbitrary gravitational field, we find
\begin{equation}\label{A7}
\left[\begin{array}{cc}
\mathcal {E} & \mathcal{B}\cr
\mathcal{B}^{\rm T} & \mathcal{S}\cr 
\end{array}\right]\,,
\end{equation}
where $\mathcal{E}$ and $\mathcal{S}$ are symmetric $3\times 3$ matrices and $\mathcal{B}$ is traceless due to the symmetries of the Riemann tensor. In the Melvin universe, we find
\begin{equation}\label{A8}
\mathcal{E} = {\rm diag}(\alpha, \alpha, \beta)\,, \qquad \mathcal{B} = 0\,,\qquad \mathcal{S} = {\rm diag}(-\alpha, \alpha, \gamma)\,,
\end{equation}
where $\mathcal{E}$, $\mathcal{B}$ and $\mathcal{S}$ denote the gravitoelectric,  gravitomagnetic and spatial components of the Riemann curvature tensor  as measured by the reference observer, respectively. From Eqs.~\eqref{I7} and~\eqref{I15}, we find
\begin{equation}\label{A9}
\alpha = \frac{2(1-\rho^2)}{a^2(1+\rho^2)^4}\,, \quad \beta =  \frac{4\rho^2}{a^2(1+\rho^2)^4}\,, \quad \gamma= \frac{4 \rho^2(2-\rho^2)}{a^2(1+\rho^2)^8}\,.
\end{equation}

We can now express the metric of Melvin's magnetic universe in terms of local Fermi coordinates. For the sake of convenience, we henceforth express Fermi coordinates as $X^{\hat \mu} = (T, X, Y, Z)$, where the spatial origin of this coordinate system is permanently occupied by the fiducial observer.  The Fermi metric is
\begin{align}\label{A10}
ds^2 ={}& - (1+ 2 \hat{\Phi}) dT^2 + dX^2 + dY^2+ dZ^2    \nonumber   \\
{}& +\frac{1}{3} \alpha (YdZ-ZdY)^2 - \frac{1}{3} \alpha (ZdX-XdZ)^2 - \frac{1}{3} \gamma (XdY-YdX)^2\,,
\end{align}
where the gravitoelectric  potential $\hat{\Phi}$ is given by
\begin{equation}\label{A11}
\hat{\Phi} = \frac{2\rho X}{a(1+\rho^2)^2}+ \frac{1}{a^2(1+\rho^2)^4}[X^2 + Y^2 +\rho^2 (X^2 - Y^2 + 2 Z^2)]\,.
\end{equation}
Therefore, in this approximation scheme,  the gravitoelectric  potential is given by $gX$ plus the lowest-order tidal terms that are quadratic in spatial Fermi coordinates. For the fiducial observer at the symmetry axis ($\rho = 0$), 
$\hat{\Phi} = (X^2 + Y^2)/a^2$, while for the corresponding observer at $\rho = 1$, $\hat{\Phi} =  X/(2a) + (X^2 + Z^2)/(8a^2)$.

\section{Circular Orbits of Charged Particles}

We define circular orbits to be those with constant radial coordinate $\rho$. The equation of motion of a charged particle is
\beq
\label{B1}
\frac{d^2x^\mu}{d\tau^2}+\Gamma^\mu{}_{\alpha\beta}\frac{dx^\alpha}{d\tau} \frac{dx^\beta}{d\tau}= \frac{q}{m} F^\mu{}_{\nu}\frac{dx^\nu}{d\tau}\,,
\eeq
which for the constant radial coordinate $\hat{\rho}_0$ reduces to 
\beq
\label{B2}
\left(\frac{dt}{d\tau}\right)^2 - \left(\frac{dz}{d\tau}\right)^2 = \frac{1}{2}\frac{1-\hat{\rho}_0^2}{(1+\hat{\rho}_0^2)^4} \left(\frac{d\phi}{d\tau}\right)^2+ \frac{1}{a}\frac{\hat{q}}{(1+\hat{\rho}_0^2)^3} \frac{d\phi}{d\tau}\,,
\eeq
where $\hat{q} = q/m$. Moreover, the tangent to the world line of the particle is a unit timelike vector; therefore, 
\beq
\label{B3}
\left(\frac{dt}{d\tau}\right)^2 - \left(\frac{dz}{d\tau}\right)^2 = \frac{1}{a^2(1+\hat{\rho}_0^2)^2} + \frac{\hat{\rho}_0^2}{(1+\hat{\rho}_0^2)^4} \left(\frac{d\phi}{d\tau}\right)^2\,.
\eeq
Equations~\eqref{B2} and~\eqref{B3} together imply
\beq
\label{B4}
\left(\frac{d\phi}{d\tau} +\frac{\hat{q}}{a} \frac{1+\hat{\rho}_0^2}{1-3\hat{\rho}_0^2}\right)^2 = \frac{2}{a^2} \left(\frac{1+\hat{\rho}_0^2}{1-3\hat{\rho}_0^2}\right)^2(1-3\hat{\rho}_0^2 +\tfrac{1}{2}\hat{q}^2)\,.
\eeq
It follows from this relation that we have timelike circular orbits for
\beq
\label{B5}
1-3\hat{\rho}_0^2 +\frac{1}{2}\hat{q}^2 > 0\,.
\eeq

Using the above results, it is straightforward to compute $(\hat{P}_\phi)_0$ and $\hat{\mathbb{E}}^2_0$ for a charged particle with circular orbit of constant radius $\hat{\rho}_0$ given by the unique real solution of Eq.~\eqref{M8}. After some algebra, we find, 
\beq
\label{B6}
(\hat{P}_\phi)_0= -a \hat{q} \frac{1-2\hat{\rho}_0^2}{(1+\hat{\rho}_0^2)(1-3\hat{\rho}_0^2)} \pm \sqrt{2} a \hat{\rho}_0^2 \frac{(1 +\tfrac{1}{2}\hat{q}^2 -3\hat{\rho}_0^2)^{1/2}}{(1+\hat{\rho}_0^2)|1-3\hat{\rho}_0^2|}\,,
\eeq
\beq
\label{B7}
\hat{\mathbb{E}}_0^2 = a^2(1+\hat{\rho}_0^2)^2 + \frac{(1+\hat{\rho}_0^2)^2}{\hat{\rho}_0^2}\left[a \hat{q} + (1+\hat{\rho}_0^2)(\hat{P}_\phi)_0\right]^2\,,
\eeq
which reduce to Eq.~\eqref{T4} in the absence of electric charge,  as expected.

\section{Conformal Collapse: $\chi \ge 0$ for $t_* \le \sqrt{21}/4$}

Let us start with the energy-momentum tensor of the dynamic Melvin universe in the conformal collapse scenario $\mathbb{T}_{\mu \nu}$ given in Eqs.~\eqref{G5}--\eqref{G6} and define three manifestly positive quantitiess
\beq
\label{C1}
\sigma_1 := \frac{4}{(1+\rho^2)^2}\,, \qquad \sigma_2 := \frac{b^2}{\mathbb{A}^2}\,, \qquad \sigma_3 := \frac{4 b\rho}{\mathbb{A}(1+\rho^2)}\,, 
\eeq
in terms of which we can write
\beq
\label{C2}
8 \pi\,\mathbb{T}_{tt} = \sigma_1 + 3\, \sigma_2\,, \qquad 8 \pi\,\mathbb{T}_{\rho \rho} = \sigma_1 +  \sigma_2\,, \qquad 8 \pi\,\mathbb{T}_{zz} =  -\sigma_1 + \sigma_2\,, 
\eeq
\beq
\label{C3}
8 \pi\,\mathbb{T}_{\phi \phi} = \frac{\rho^2}{(1+\rho^2)^4}(\sigma_1 + \sigma_2)\,, \qquad 8 \pi\,\mathbb{T}_{t \rho} = 8 \pi\,\mathbb{T}_{\rho t} = -\sigma_3\,. 
\eeq

We consider an arbitrary timelike or null vector $W^\alpha$ and note that $\tilde{g}_{\mu \nu}W^\mu W^\nu \le 0$ implies
\beq
\label{C4}
(W^t)^2 - (W^z)^2 \ge (W^\rho)^2  +\frac{\rho^2}{(1+\rho^2)^4}(W^\phi)^2\,. 
\eeq
We are interested in the quantity $8 \pi \chi = 8 \pi \mathbb{T}_{\mu \nu} W^\mu W^\nu$, which can be written as 
\begin{align}
\label{C5}
\nonumber 8 \pi \chi = {}&(\sigma_1 + 3\,\sigma_2)(W^t)^2 - 2 \sigma_3 W^t W^\rho + (-\sigma_1 + \sigma_2) (W^z)^2 \\
{}&+ (\sigma_1 + \sigma_2)(W^\rho)^2  +\frac{\rho^2}{(1+\rho^2)^4}(\sigma_1 + \sigma_2)(W^\phi)^2\,. 
\end{align}

Let us first note that $\chi$ can be negative. For instance, in the case of a radial null geodesic, we have $W^t = W^\rho$, $W^\phi = W^z = 0$ and the magnitude of $\chi$ is then simply proportional to $\sigma_1 +2 \sigma_2-\sigma_3$. For $\rho = 3$ and $\mathbb{A}/b = 2$, we have $\sigma_1 = 1/25$,  $\sigma_2 = 1/4$ and $\sigma_3 = 3/5$; hence, $\sigma_1 +2 \sigma_2-\sigma_3 = -3/50$ and $\chi < 0$. 

It is convenient to write $\chi = \chi_1 + \chi_2 + \chi_3$, where
\beq
\label{C6}
 8 \pi \chi_1 = \sigma_1 (W^t)^2 -\sigma_1  (W^z)^2\,, 
\eeq
\beq
\label{C7}
 8 \pi \chi_2 =  \sigma_2 (W^z)^2  +\frac{\rho^2}{(1+\rho^2)^4}(\sigma_1 + \sigma_2)(W^\phi)^2\,
\eeq
and
\beq
\label{C8}
 8 \pi \chi_3 =  3\,\sigma_2(W^t)^2 - 2 \sigma_3 W^t W^\rho + (\sigma_1 + \sigma_2)(W^\rho)^2\,. 
\eeq
Here, $\chi_1$ is positive via Eq.~\eqref{C4}, while $\chi_2$ is a sum of manifestly positive quantities. For $\chi_3$, let us use Eq.~\eqref{C1} and note that 
\beq
\label{C9}
 8 \pi \chi_3 =  3\,\left(\frac{b}{\mathbb{A}}W^t -\frac{4}{3}\frac{\rho}{1+\rho^2} W^\rho\right)^2 + 4\,\left[ \frac{b^2}{4\mathbb{A}^2} - \frac{\tfrac{4}{3}\rho^2-1}{(1+\rho^2)^2}\right] (W^\rho)^2\,. 
\eeq
Therefore, $\chi \ge 0$ if we ensure that
\beq
\label{C10}
\frac{b^2}{4\mathbb{A}^2} \ge \frac{\tfrac{4}{3}\rho^2-1}{(1+\rho^2)^2}\,. 
\eeq
This relation has a function of $\rho$ on its right-hand side that starts from a minimum value of $-1$ at $\rho = 0$, increases with increasing $\rho$ and has a maximum value of $4/21$ at $\rho_{\rm max} = \sqrt{5/2}$; then, it decreases monotonically and  vanishes asymptotically as $\rho \to \infty$. Hence, Eq.~\eqref{C10} is satisfied provided $b^2/\mathbb{A}^2 \ge 16/21$ or $\mathbb{A}/b \le \sqrt{21}/4$. We know that $\mathbb{A}$ decreases from $a$ at $t=0$ and vanishes at $t_* = a/b$; hence, we must have $t_* = a/b \le \sqrt{21}/4$. In this way, $\chi_3$ is positive and the weak and strong energy conditions are satisfied.

\section{Axial Collapse Scenario}

There are different ways that we can consider the collapse of the Melvin universe in an axial manner. For instance, one can modify  Melvin metric~\eqref{I4} and assume radial collapse toward the axis
\beq  \label{D1}
ds^2=a^2 (1+\rho^2)^2(-dt^2+dz^2)+ a^2 h^2(t) (1+\rho^2)^2 d\rho^2 + a^2 \frac{\rho^2}{(1+\rho^2)^2}d\phi^2\,
\eeq   
together with the electromagnetic potential $A = -a h(t)(1+\rho^2)^{-1} d\phi$; then, Maxwell's equations are satisfied provided $h(t) = (1 - b' t)^{1/2}$, where $b' > 0$ is a constant.  

Another natural way would be to assume that the whole horizontal plane collapses, namely, 
\beq   \label{D2}
ds^2=a^2 (1+\rho^2)^2(-dt^2+dz^2)+ a^2 H^2(t) \left[(1+\rho^2)^2 d\rho^2 +  \frac{\rho^2}{(1+\rho^2)^2}d\phi^2 \right]\,,
\eeq
where we assume 
\beq  \label{D3}
A =  -\frac{a H(t)}{1+\rho^2} d\phi\,.
\eeq
Hence, 
\beq   \label{D3a}
F = dA= -\frac{a}{1+\rho^2}\frac{dH}{dt} dt\wedge d\phi + \frac{2 a\rho H(t)}{(1+\rho^2)^2} d\rho \wedge d\phi\,.
\eeq
Maxwell's equations are satisfied in this case with
\beq  \label{D4}
H(t) = 1 - \kappa\, t\,,
\eeq
where $\kappa > 0$ is a dimensionless constant.  Remarkably, we have the same kind of linear temporal dependence as in Section V; therefore, we proceed with the investigation of this case. The fiducial observers at rest 
in space carry the tetrad frame
\beq
\label{D5}
e_{\hat 0}=\frac{1}{a(1+\rho^2)}\partial_t\,,\quad e_{\hat 1}=\frac{1}{a H(t)(1+\rho^2)}\partial_\rho\,,\quad
e_{\hat 2}=\frac{1+\rho^2}{a H(t)\rho}\partial_\phi\,,\quad e_{\hat 3}=\frac{1}{a(1+\rho^2)}\partial_z\,.
\eeq
It follows from Eq.~\eqref{D3} that there is a vertical magnetic field and an azimuthal electric field that diverges on the symmetry axis given by
\beq  \label{D6}
\mathfrak{B} = \frac{2}{Ha (1+\rho^2)^2}\,,\qquad \mathfrak{E} = \frac{\kappa}{Ha\rho(1+\rho^2)}\,.
\eeq
The corresponding electromagnetic invariants are $^\kappa I_1=  \mathfrak{B}^2 - \mathfrak{E}^2$ and $^\kappa I_2=0$.

The acceleration tensor of the fiducial observers in this case is
\beq    \label{D7}
^\kappa\mathbf{g} =\frac{2\rho}{a H(t)(1+\rho^2)^2}\,e_{\hat1}\,,\qquad ^\kappa\boldsymbol{\omega} =0\,.
\eeq
Thus, as before, the observer is attracted to the symmetry axis and this attraction tends to infinity as $H(t) \to 0$.

The complete energy-momentum source of the gravitational field, $^\kappa\mathbb {T}^\mu{}_\nu$, in this case has diagonal  components given by
\beq    \label{D8}
8\pi\, ^\kappa\mathbb{T}^\mu{}_\nu = \frac{4}{a^2H^2 (1+\rho^2)^4}{\rm diag}[-1,1,1,-1]-\frac{\kappa^2}{a^2H^{2}(1+\rho^2)^2} {\rm diag}[1,0,0,1]
\eeq
and off-diagonal components 
\beq    \label{D9}
8\pi\, ^\kappa\mathbb{T}^t{}_\rho=\frac{4\rho \kappa}{  a^2 H (1+\rho^2)^3} \,,\qquad 8\pi \,^\kappa\mathbb {T}^\rho{}_t=-\frac{4\rho\kappa }{  a^2 H^3 (1+\rho^2)^3}\,.
\eeq
 In this case, $^\kappa\chi := \,^\kappa\mathbb{T}_{\mu \nu} W^\mu W^\nu$ turns out to be negative for an \emph{outgoing} radial null vector for $\rho = 3$ and $\kappa = 0.1$; therefore, the weak energy condition is violated, as expected. Moreover, the Kretschmann scalar is given by
\beq   \label{D10}
^\kappa K = \frac{4}{a^4H^4 (1+\rho^2)^4}\left[ \kappa^4 -\frac{8\kappa^2 (3\rho^2-2)}{(1+\rho^2)^2}+\frac{16(3\rho^4-6\rho^2+5)}{(1+\rho^2)^4}\right] \,.
\eeq
The physical components of the energy-momentum tensor, as measured by the fiducial observers, as well as the Kretschmann scalar are all finite in space and diverge in time when $H = 0$, which is the essential singularity of this spacetime. 

The geodesic equations have been numerically integrated in this case and our numerical experiments confirm the formation of the cosmic double-jet pattern along the axis of symmetry; that is, we find that, as expected,  future-directed timelike geodesic paths are attracted toward the symmetry axis and their Lorentz factors, as measured by the fiducial observers at rest in space, approach infinity as $H \to 0$. 

To extend our considerations to the motion of charged particles, we follow the same approach to the Lorentz force equation as in the last part of Section VI. The existence of the Killing vectors $\partial_\phi$ and $\partial_z$ in this dynamic spacetime implies 
\beq    \label{D11}
\frac{d\phi}{d\tau} =\frac{1+\rho^2}{a^2 H^2\rho^2}\,[a\hat{q}H +(1+\rho^2)\,^{\kappa}P_\phi]\,, \quad  a^2 (1+\rho^2)^2 \frac{dz}{d\tau} =\, ^{\kappa}P_z\,,
\eeq
where $^{\kappa}P_\phi$ and $^{\kappa}P_z$ are constants of the motion. Moreover, from $U^\mu U_\mu = -1$,  where $U^\mu= dx^\mu /d\tau$ is the 4-velocity of the charged particle, we find
\beq     \label{D12}
\left(\frac{dt}{d\tau}\right)^2 - H^2\left(\frac{d\rho}{d\tau}\right)^2 = \frac{1}{a^2 (1+ \rho^2 )^2} + \frac{^{\kappa}P_z^2}{a^4 (1+ \rho^2 )^4}+ \frac{[(1+\rho^2)^{\kappa}P_\phi+ a\hat{q} H]^2}{a^4H^2 \rho^2(1+ \rho^2 )^2}\,.
\eeq
To find an equation for $\rho(\tau)$, we consider the radial component of the Lorentz force equation
\beq
\label{D13}
\frac{dU^1}{d\tau}+\Gamma^1{}_{\alpha\beta}U^\alpha U^\beta=  \hat{q} F^1{}_{\nu}U^\nu\,, 
\eeq
which in this case implies
\begin{align}
\label{D14}
\nonumber \frac{d^2\rho}{d\tau^2}{}&+\frac{2\rho}{1+\rho^2}\left[\frac{1}{H^2} \left(\frac{dt}{d\tau}\right)^2 + \left(\frac{d\rho}{d\tau}\right)^2- \frac{1}{H^2}\left(\frac{dz}{d\tau}\right)^2\right] - 2\frac{\kappa}{H(t)}\frac{dt}{d\tau}\frac{d\rho}{d\tau} \\
{}& -\rho\, \frac{1-\rho^2}{(1+\rho^2)^5} \left(\frac{d\phi}{d\tau}\right)^2 = 2 \hat{q} \frac{\rho}{a H(t) (1+ \rho^2 )^4}\frac{d\phi}{d\tau}\,.
\end{align}
With proper boundary conditions, the above equations can be integrated to determine the motion of charged particles in this dynamic Melvin spacetime.

We have studied the motion of charged particles in this case via numerical integration of the Lorentz force equation. The results of our numerical investigations are qualitatively the same as in the conformal collapse scenario outlined at the end of Section VI and are consistent with the formation of cosmic double-jet configurations.

\end{document}